\documentclass[12pt,onecolumn,draftcls]{IEEEtran}
\IEEEoverridecommandlockouts
\usepackage{amsmath}
\usepackage{amsfonts}
\usepackage{bbding}
\usepackage{amssymb}
\usepackage{array}
\usepackage{subfigure}

\usepackage{graphicx}
\usepackage{subfigure}
\usepackage[named]{algo}
\usepackage{algorithmic}
\usepackage{algorithm}

\usepackage{psfrag}
\usepackage{xfrac}
\usepackage{stfloats}
\usepackage[compress]{cite}
\makeatletter
\renewcommand{\citepunct}{,\penalty\@m\hskip.13emplus.1emminus.1em}
\renewcommand{\citedash}{\hbox{--}\penalty\@m}
\makeatother
\usepackage{setspace}
\usepackage{color}
\usepackage{xcolor}
\allowdisplaybreaks

\usepackage{amsthm}
\usepackage{stfloats}

\newtheorem{rem}{Remark}

\newtheorem{prop}{Proposition}

\usepackage{hyperref}
\usepackage{bm}

\begin{document}
\title{Unsupervised Deep Learning for Optimizing Wireless Systems with Instantaneous and Statistic Constraints}

\author{
\IEEEauthorblockN{{Chengjian Sun, Changyang She and Chenyang Yang}} \vspace{0.0cm}

\thanks{This paper has been presented in part at the IEEE Global Communications Conference 2019 \cite{sun2019unsupervised} and in part at the IEEE International Symposium on Personal, Indoor and Mobile Radio Communications 2019 \cite{sun2019PIMRC}.}

\thanks{C. Sun and C. Yang are with the School of Electronics and Information Engineering, Beihang University, Beijing 100191, China (email:\{sunchengjian,cyyang\}@buaa.edu.cn).}

\thanks{C. She is with the School of Electrical and Information Engineering, University of Sydney, Sydney, NSW 2006, Australia (e-mail: shechangyang@gmail.com).}
}
\maketitle
\begin{abstract}
Deep neural networks (DNNs) have been introduced for designing wireless policies by approximating the mappings from environmental parameters to solutions of optimization problems. Considering that labeled training samples are hard to obtain, unsupervised deep learning has been proposed to solve functional optimization problems with statistical constraints recently. However, most existing problems in wireless communications are variable optimizations, and many problems are with instantaneous constraints. In this paper, we establish a unified framework of using unsupervised deep learning to solve both kinds of problems with both instantaneous and statistic constraints. For a constrained variable optimization, we first convert it into an equivalent functional optimization problem with instantaneous constraints. Then, to ensure the instantaneous constraints in the functional optimization problems, we use DNN to approximate the Lagrange multiplier functions, which is trained together with a DNN to approximate the policy.
We take two resource allocation problems in ultra-reliable and low-latency communications as examples to illustrate how to guarantee the complex and stringent quality-of-service (QoS) constraints with the framework. Simulation results show that unsupervised learning outperforms supervised learning in terms of QoS violation probability and approximation accuracy of the optimal policy, and can converge rapidly with pre-training.

\end{abstract}
\begin{IEEEkeywords}
Unsupervised deep learning, variable optimization, functional optimization, constraints, ultra-reliable and low-latency communications
\end{IEEEkeywords}

\section{Introduction}
Beyond fifth generation (B5G) cellular systems are expected to support diverse quality-of-service (QoS) requirements of various applications, say video streaming and ultra-reliable and low-latency communications (URLLC)  \cite{3GPP2017Agree}. To efficiently use network resources to satisfy the QoS requirements in a dynamic environment, a base station (BS) needs to optimize its transmission policy  according to the environment parameters before they change. A typical wireless policy adapts to small-scale channels, e.g., power allocation, beamforming, and user scheduling, where the BS needs to find the optimal policy every few milliseconds, depending on the channel coherence time. If the policy cannot be obtained in closed-form, which is the case for many problems in wireless communications, numerical algorithms (e.g., interior-point method) have to be used for finding the solution. This incurs high computational overheads. If the computing time used for searching the optimal solution is longer than the channel coherence time, the obtained solution cannot guarantee the QoS with the current channel realization. This issue becomes more critical for URLLC with 1 ms end-to-end (E2E) latency \cite{she2020deep}.


To avoid executing traditional numerical algorithms repeatedly whenever the environment status changes, a novel idea of ``learning to optimize" was proposed in  \cite{Sun2018TSP}, which finds a mapping from the environmental parameters to the optimal decision by approximating the mapping with a deep neural network (DNN). While promising, the method proposed therein needs a large number of labels to train the DNN, which are obtained by finding the solutions of the original optimization problem for given realizations of random environment parameters. This is possible if the original problem is a variable optimization problem, which aims to find a scalar (e.g., transmit power) or a vector (e.g., beamforming vector), but is very hard if not impossible when the original problem should be formulated as functional optimization problem \cite{gregory2018constrained}.

Most of existing problems in wireless communications are formulated as constrained variable optimization problems, e.g., finding beamforming vectors to maximize the sum-rate or minimize the total power subjects to the maximal transmit power constraint and the QoS constraint such as the data rate or the signal-to-noise ratio (SNR) exceeding a threshold. In these variable optimization problems, the objective function, constraints, and the variables to be optimized change in the same timescale. If they change in different timescales, the problems turn out to be functional optimization problems \cite{liu2020optimizing}.

An optimization problem that finds a function to maximize or minimize an objective belongs to functional optimization problems \cite{gregory2018constrained}, which are quite common for optimal control problems but are less familiar to the wireless community. When the timescale of the performance metric concerned by a wireless policy is much longer than the timescale of the environment parameters that the optimization depends on or the policy itself is in multi-timescale, which is often the case in cross-layer design, the policy should be found from a functional optimization problem. In general, the solutions of functional optimization problems can hardly be obtained in closed-form, which are usually obtained numerically.
One of the widely applied numerical methods for solving functional optimization problems is finite element method (FEM) \cite{Zienkiewicz1977FEM}. As a mesh-based method, FEM suffers from the curse of dimensionality, especially in the multi-user scenarios in wireless networks, where the dimension increases with the number of users. To overcome the difficulty in generating labels for training, unsupervised learning approach has been proposed to solve functional optimization problems with statistical constraints recently \cite{eisen2019learning}.

\subsection{Related Works}
Two branches of deep learning techniques have been proposed to solve the wireless optimization problems: supervised deep learning \cite{Sun2018TSP, Liu2018Lrn2Opt,GY2018,dong2020deep} and unsupervised deep learning \cite{eisen2019learning,lee2019deep,liu2020optimizing}.

The idea of ``learning to optimize"  was first proposed for variable optimization problems by the authors in  \cite{Sun2018TSP}, where approximated solutions were proved able to be obtained from fully-connected DNNs. A deep learning framework was proposed in \cite{Liu2018Lrn2Opt} to find the relationship between flow information and link usage by learning from past computation experience. To learn the optimal predictive resource allocation under the QoS constraint of video streaming, a DNN was designed and active learning was used to decrease the required labels in \cite{GY2018}. To improve the approximation accuracy, a cascaded neural network was introduced to approximate optimal resource allocation policies and deep transfer learning was applied to fine-tune the DNN in non-stationary wireless networks \cite{dong2020deep}. By training the DNNs offline, an approximated decision can be obtained with low complexity online \cite{Sun2018TSP, GY2018, Liu2018Lrn2Opt,dong2020deep}, say about $1$\% of the original numerical optimization \cite{GY2018}. Such an idea can be regarded as a kind of computing offloading over time, which shifts the computations from online to offline.  However, two issues remain to resolve in this approach of supervised learning: 1) the labels may be obtained in unaffordable complexity, which is especially true for functional optimization problems, and 2) the QoS violations caused by the approximation errors are not controlled, which makes the approach inapplicable for the wireless systems requiring stringent QoS such as URLLC.


To find the optimal policy without labeled training samples, unsupervised deep learning was introduced in \cite{YW2019} to solve a variable optimization problem with maximal power constraint and proposed in \cite{eisen2019learning,lee2019deep} to solve functional optimization problems subject to statistic constraints. In \cite{YW2019}, the previous two issues are circumvented by using the empirical average of the objective function of the optimization problem as the cost function for training a DNN and by selecting a proper activation function in the output layer of the DNN. However, whether or not using the empirically averaged objective function as the cost function can give rise to the optimal solution was not explained, and using activation function can only satisfy simple constraints such as maximum or non-negative resource constraints. In \cite{eisen2019learning}, the primal-dual method was applied to maximize and minimize the Lagrangian function of the constrained optimization problem in the primal domain and the dual domain, respectively. In the primal domain, the optimal policy is approximated by a DNN. The parameters of the DNN and the Lagrangian multipliers, i.e., the optimization variable in the dual domain, are updated iteratively. The same method was introduced to solve distributed optimization in \cite{lee2019deep}. By considering the original problems in its dual domain, complex constraints can be satisfied. However, the proposed method in \cite{eisen2019learning, lee2019deep} is only applicable to the functional optimization problems with statistical constraints.

When optimizing transmission policies in wireless communications, there exist both variable optimizations and functional optimizations, and exist both instantaneous constraints and statistic constraints. A theoretic interpretation of why variable optimization problems can be learned without supervision remains unclear. While some resource constraints can be satisfied by choosing proper activation functions \cite{eisen2019learning}, many (especially QoS) constraints are complex and hence cannot be satisfied by activation functions. How to guarantee instantaneous constraints with unsupervised deep learning remains an open problem.

%
%

\subsection{Motivation and Contributions}
In this paper, we investigate how to establish a unified framework for learning to optimize both variable and functional optimizations subject to both instantaneous and statistic constraints, and for solving functional optimizations subject to both types of constraints with unsupervised deep learning.

Since the QoS requirement in URLLC is complex and stringent, we take a downlink (DL) URLLC system as an example to show how to apply the proposed framework. In particular, we formulate two resource allocation problems with delay and reliability constraint. One is variable optimization, where a BS allocates bandwidth according to large-scale channel gains. Another is a hybrid variable and functional optimization with both instantaneous and statistic constraints, where a BS jointly allocates bandwidth according to large-scale channel gains and transmit power according to small-scale channel gains.
The main contributions are summarized as follows.
\begin{itemize}
\item We prove that the mapping from environment parameters to the solution of a constrained variable optimization problem can be formulated as a proper functional optimization problem with instantaneous constraints. Then, we develop a unified framework for using unsupervised deep learning to find the approximated optimal policy from both variable and functional optimization problems. Different from the method in \cite{eisen2019learning,lee2019deep} that only considers statistic constraints, both instantaneous and statistic constraints are considered in our framework.
\item We illustrate how to solve functional optimization problems with the bandwidth and power allocation problem in URLLC. We derive global optimal solution of the problem from its first-order necessary conditions in a symmetric scenario, where the QoS requirements, packet arrival rates, and large-scale channel gains of all users are identical.
 Simulation and numerical results show that performance achieved by the unsupervised learning is very close to that of the optimal policy, and is superior to supervised deep learning in terms of QoS guarantee and the policy approximation accuracy.
\end{itemize}

The rest of the paper is organized as follows. In Section II, we show how to convert a variable optimization problem into a functional optimization problem and how to solve functional optimization problems subject to both instantaneous and statistic constraints with unsupervised deep learning. In Section III, we consider two resource allocation problems in URLLC systems to illustrate how to use the proposed framework. 
Simulation and numerical results are provided in Section IV. We conclude this paper in Section V.

\section{Unsupervised Deep Learning for Variable and Functional Optimizations}
In this section, we first introduce the definitions of functional and functional optimization. Then, we {prove that a constrained continuous variable optimization problem can be equivalently converted into a functional optimization problem with instantaneous constraints.} Next, we {introduce} functional optimization problem {with statistic constraint} in wireless networks by an example, the classical water-filling power control. Finally, we present a framework to solve functional optimization problems with both instantaneous and statistic constraints using unsupervised deep learning.

\subsection{Functional and Functional Optimization}
According to the definition in  \cite{Calculus2012Daniel}, a \emph{functional} is a function of a function, which maps a function into a scalar. Functional is a kind of functions, where the ``variable'' itself is a function. A general type of functionals can be expressed as an integral of functions, say
\begin{align}
\mathcal{F}[\bm{x}(\bm{\theta})] = \int_{\bm{\theta} \in \mathcal{D}_\theta} {F_{0}\left[\bm{x}(\bm{\theta});\bm{\theta}\right] \mathrm{d} \bm{\theta}},\nonumber
\end{align}
where $\mathcal{F}[\bm{x}(\bm{\theta})]$ is a functional since its ``variable''  $\bm{x}(\bm{\theta})$ is a function of $\bm{\theta}$, and
$F_{0}\left[\bm{x}(\bm{\theta});\bm{\theta}\right]$ is a function of two group of variables, a specific value of $\bm{\theta}$ and the corresponding value of $\bm{x}(\bm{\theta})$.

An optimization problem is a \emph{functional optimization problem} if either the objective function or the constraint is a functional.

\subsection{Functional Optimization Problem with Instantaneous Constraints}\label{subsec:func}
Consider a continuous variable optimization problem that finds a vector $\bm{x} \!\in\! \mathcal{D}_x \!\subseteq\! \mathbb{R}^{N_x}$ consisting of $N_x$ variables to minimize objective $f\left(\bm{x};\bm{\theta}\right)$ under constraints ${C}_i\left(\bm{x};\bm{\theta}\right)$,
\begin{align}   \label{prob:VarOpt}
    \mathop \mathrm{min} \limits_{\bm{x}} \quad& f\left(\bm{x};\bm{\theta}\right) \\
    \text{s.t.} \quad & {C}_i\left(\bm{x};\bm{\theta}\right) \leq 0, i = 1,...,I, \label{con} \tag{\theequation a}
\end{align}
where $\bm{\theta} \!\in\! \mathcal{D}_\theta \!\subseteq\! \mathbb{R}^{N_\theta}$ is a vector of $N_\theta$ environmental parameters, which is a realization of continuous random variables and is assumed known for optimization, $\mathcal{D}_\theta$ is a compact set, and $f\left(\bm{x};\bm{\theta}\right)$ and ${C}_i\left(\bm{x};\bm{\theta}\right)$
are differentiable with respect to (w.r.t.) $\bm{x}$ and $\bm{\theta}$.
Since the constraint $\bm{x} \!\in\! \mathcal{D}_x$ can be considered as a special case of \eqref{con}, it is not listed explicitly.

For example, $\bm{x}$ is a beamforming vector, and $\bm{\theta}$ is a channel vector that is known by estimation at the BS before optimizing beamforming. Another example is the predictive resource allocation problem in \cite{GY2018}, where $\bm{x}$ is a matrix composing of the fractions of bandwidth assigned to several mobile users in the frames of a prediction window, and $\bm{\theta}$ is a matrix consisting of future average data rates in the frames of these users that are known by prediction before the optimization.
In most of the cases, the closed-form optimal solution of problem \eqref{prob:VarOpt} can hardly be obtained from the Karush-Kuhn-Tucker (KKT) conditions. As a result, one needs to search for the optimal solution numerically again whenever the value of $\bm{\theta}$ changes and hence needs to be updated by estimation or prediction. For the example of beamforming, the update duration is the channel coherence time. For the example in \cite{GY2018}, the update duration is the duration of the prediction window, within which the large scale channel gains (and hence the average data rates) may change among frames. To facilitate practical use for wireless applications with fast changing environmental parameters,
a promising approach is to find the mapping from $\bm{\theta}$ to the optimal solution, i.e., find the function $\bm{x}^*(\bm{\theta})$. This can be obtained by supervised learning, where a DNN is used to approximate $\bm{x}^*(\bm{\theta})$ and is trained with the labels generated by solving problem \eqref{prob:VarOpt} for a large number of realizations of $\bm{\theta}$ \cite{Sun2018TSP}.

To avoid generating labels by solving a variable optimization problem,
one can resort to unsupervised deep learning by using the objective function of the problem as the loss function for training the DNN. Yet this is not straightforward since the objective function in \eqref{prob:VarOpt} is a function of $\bm{x}$ and $\bm{\theta}$, rather than a function of \emph{the function to be optimized, i.e., $\bm{x}(\bm{\theta})$}.

In fact, the mapping from the environmental parameters to the optimal solution of problem \eqref{prob:VarOpt} can be found from a functional optimization problem.
Then, the issue becomes: how to formulate such a functional optimization problem?

In order to find the function $\bm{x}^*(\bm{\theta})$, we construct the following functional optimization problem,
\begin{align}   \label{prob:FuncOpt}
    \mathop \mathrm{min} \limits_{\bm{x}(\bm{\theta})} \quad& \mathbb{E}_{\bm{\theta}} \left\{f\left[\bm{x}(\bm{\theta});\bm{\theta}\right] \right\}=\int_{\bm{\theta} \in \mathcal{D}_\theta} {f\left[\bm{x}(\bm{\theta});\bm{\theta}\right] p(\bm{\theta}) \mathrm{d} \bm{\theta}} \\
    \text{s.t.} \quad & {C}_i\left[\bm{x}(\bm{\theta});\bm{\theta}\right] \leq 0, \  \forall {\bm{\theta} \in \mathcal{D}_{\theta}},  i = 1,...,I, \label{con2} \tag{\theequation a}
\end{align}
where $\bm{x}(\bm{\theta})$ is optimized to minimize the expectation of the objective function in problem \eqref{prob:VarOpt} over $\bm{\theta}$, and $p(\bm{\theta})$ is the probability density function (PDF) of $\bm{\theta}$. This is a functional optimization problem since the objective function in \eqref{prob:FuncOpt} is a function of the function $\bm{x}(\bm{\theta})$.

The constraints in problems \eqref{prob:VarOpt} and \eqref{prob:FuncOpt} are not functionals, because the left-hand sides of them {only} depend on \emph{the realizations of the random environment parameters $\bm{\theta}$}.  We refer to this kind of constraints as {\bf instantaneous constraints}. For example, when the beamforming vector is optimized according to the channel vector known at a BS, the instantaneous data rate constraint or the transmit power constraint belongs to the instantaneous constraints.

It is worth noting that the constraints in the two problems are different. The constraints in \eqref{con} needs to be ensured for a {specific} realization of $\bm{\theta}$. As a result, the solution of problem \eqref{prob:VarOpt} is optimal only for the given realization of $\bm{\theta}$. Once the value of $\bm{\theta}$ varies, the problem needs to be solved again. However, the constraints in \eqref{con2} should be satisfied for all the possible realizations of $\bm{\theta} \in \mathcal{D}_{\theta}$. Therefore, the solution of problem \eqref{prob:FuncOpt}, denoted by $\bm{x}_{\rm opt}(\bm{\theta})$, is optimal for arbitrary realization of $\bm{\theta}$. When the environment status changes, {the optimal solution can be immediately obtained from $\bm{x}_{\rm opt}(\bm{\theta})$,} and there is no need to solve the problem again. 

\begin{prop}\label{P:1}
\emph{$\bm{x}^*(\bm{\theta})$ is optimal for problem \eqref{prob:FuncOpt}, and the value of $\bm{x}_{\rm opt}(\bm{\theta})$ given arbitrary realization of $\bm{\theta}$ is optimal for problem \eqref{prob:VarOpt}  with probability one.}
\end{prop}

This proposition is proved in Appendix~\ref{App:Upgrade}. It indicates that a constrained continuous variable optimization problem can be equivalently converted into a functional optimization problem with instantaneous constraints in the sense of almost surely finding the same mapping.


\subsection{Functional Optimization Problem with Statistic Constraints}
If the timescale in a wireless application for measuring the system performance or the QoS is much longer than the update duration of the environment parameters for the optimization, or the timescales of the ``variables" to be optimized differ, then the objective function or the constraint will be a functional.
To help understand, we re-visit the classic power control problem \cite{WirelessCom}, which adjusts transmit power $P(g)$ according to small-scale channel gain $g$. The goal is to maximize the ergodic capacity subject to the average transmit power constraint,
\begin{align}
\mathop {\max }\limits_{P\left( g \right)}\;&\quad \mathbb{E}_g \left\{W{{\log }_2}\left[ {1 + \frac{{\alpha gP\left( g \right)}}{{{N_0}W}}} \right] \right\} = \int_0^\infty {W{{\log }_2}\left[ {1 + \frac{{\alpha gP\left( g \right)}}{{{N_0}W}}} \right]} p\left( g \right)\mathrm{d}g,\label{eq:examp1}\\
\text{s.t.}\;& \int_0^\infty  {P\left( g \right)} p\left( g \right)\mathrm{d}g \le {P_{{\rm{ave}}}}\label{eq:Cexamp1}\tag{\theequation a},
\end{align}
where $W$ is the bandwidth, $P_{\rm ave}$ is the maximal average transmit power, $\alpha$ is the large-scale channel gain, $p(g)$ is the PDF of the small-scale channel gain, and $N_0$ is the single-side noise spectral density. This is a functional optimization problem, since both the objective and the constraint are functional, which are measured in a timescale much longer than the update duration of the environment parameter for the optimization, i.e., channel coherence time.


Unlike the instantaneous constraints in \eqref{con} and \eqref{con2}, the left-hand side of constraint in \eqref{eq:Cexamp1} relies on the \emph{distribution rather than a specific realization of the environmental parameter} $g$. We referred to the constraints depending on the distribution of $\bm{\theta}$ as {\bf statistic constraints}.

\subsection{A Framework of Solving Functional Optimization with Both Types of Constraints}\label{subsec:funcDNN}
A functional optimization problem with $I$ instantaneous constraints and $J$ statistic constraints can be expressed as follows,
\begin{align}
\mathop \mathrm{min} \limits_{\bm{x}(\bm{\theta})} \quad& {\mathbb{E}}_{\bm{\theta}} \{f\left[\bm{x}(\bm{\theta}),\bm{\theta}\right] \} \label{prob:GeneralFunc} \\
    \text{s.t.} \quad & {C}_i\left[\bm{x}(\bm{\theta}),\bm{\theta}\right] \leq 0, \  \forall {\bm{\theta} \in \mathcal{D}_{\theta}},  i = 1,...,I, \label{eq:instant} \tag{\theequation a}\\
    & {\mathbb{E}}_{\bm{\theta}}\left\{{C}_j\left[\bm{x}(\bm{\theta}),\bm{\theta}\right]\right\} \leq 0,  j = I+1,...,I+J. \label{eq:stat} \tag{\theequation b}
\end{align}

To find the optimal solution of problem \eqref{prob:GeneralFunc}, we first define the Lagrangian of the problem as
\begin{align}
L \triangleq  \int\limits_{{\bm{\theta}} \in {D_{\theta}}}{f\left[\bm{x}(\bm{\theta}),\bm{\theta}\right]p({\bm{\theta}}) {\rm{d}}{\bm{\theta}} } +\sum\limits_{i = 1}^I {\int\limits_{{\bm{\theta}} \in {D_{\theta}}} {v_i({\bm{\theta}}) {C_i\left[\bm{x}(\bm{\theta}),\bm{\theta}\right]} p({\bm{\theta}}) {\rm{d}}{\bm{\theta}}} } + \sum_{j=I+1}^{I+J}{\lambda_j{\int\limits_{{\bm{\theta}} \in {D_{\theta}}} { {C_j\left[\bm{x}(\bm{\theta}),\bm{\theta}\right]} p({\bm{\theta}}) {\rm{d}}{\bm{\theta}}} }},\nonumber
\end{align}
where ${v}_i({\bm{\theta}}) \geq 0, \forall \theta \in \mathcal{D}_{\theta}$, and ${\lambda}_j \geq 0$ are the Lagrange multipliers. Noting that every Lagrange multiplier related to each instantaneous constraint in \eqref{eq:instant} is a function of ${\bm{\theta}}$, because the constraint should be satisfied for all the possible values of $\theta$.

\subsubsection{Theoretical Approach} The theory of calculus of variations in \cite{gregory2018constrained} indicates that the optimal solution of problem \eqref{prob:GeneralFunc} should satisfy the following conditions,
\begin{align}
    & {\frac{\mathrm{\delta} L}{\mathrm{\delta} \bm{x}(\bm{\theta})} = 0,} \label{KKT_x0}\\
    & v_i({\bm{\theta}}) {C_i\left[\bm{x}(\bm{\theta}),\bm{\theta}\right]} = 0, \  \forall {\bm{\theta} \in \mathcal{D}_{\theta}}, \label{KKT_lambda} \\
    & \lambda_j{{{\mathbb{E}}_{\bm{\theta}}}\left\{ C_j\left[\bm{x}(\bm{\theta}),\bm{\theta}\right] \right\}} = 0, \label{KKT_nu} \\
    & v_i(\bm{\theta}) \geq {0}, \  \forall {\bm{\theta} \in \mathcal{D}_{\theta}}, \lambda_j \geq 0, \label{con:lambda}\\
    & \eqref{eq:instant}, \ \eqref{eq:stat}, \  \forall {\bm{\theta} \in \mathcal{D}_\theta}. \nonumber
\end{align}
From the definition of the Lagrangian, \eqref{KKT_x0} can be derived as follows,
\begin{align}
\left\{\frac{\partial{f\left[\bm{x}(\bm{\theta}),\bm{\theta}\right]}}{\partial{\bm{x}(\bm{\theta})}} + \sum\limits_{i = 1}^I { {v_i({\bm{\theta}}) \frac{\partial{C_i\left[\bm{x}(\bm{\theta}),\bm{\theta}\right]}}{\partial{\bm{x}(\bm{\theta})}}} } + \sum_{j=I+1}^{I+J}{\lambda_j{{ \frac{\partial{C_j\left[\bm{x}(\bm{\theta}),\bm{\theta}\right]}}{\partial{\bm{x}(\bm{\theta})}} } }}\right\} p({\bm{\theta}}) = \bm{0}, \  \forall {\bm{\theta} \in \mathcal{D}_{\theta}}, \label{KKT_x}
\end{align}
which is the simplified form of the Eular-Lagrange equation {defined in \cite{gregory2018constrained}}.

These conditions are the first-order necessary conditions to achieve the optimality of functional optimization problems, like the KKT conditions of variable optimization problems \cite{boyd}. However, the condition in \eqref{KKT_nu} is an integral equation, which comes from the statistic constraints. This makes solving functional optimization problems rather challenging.
In particular, even if the closed-form expression of $\bm{x}(\bm{\theta})$ can be obtained from \eqref{KKT_x0}, \eqref{KKT_lambda}, \eqref{con:lambda}, \eqref{eq:instant} and \eqref{eq:stat}, the closed-form expressions of Lagrange multiplier for the statistic constraints $\lambda_j, j=I+1,...,I+J$ are hard to derive since  integral equations are in general difficult to solve. For example, the optimal solution of problem \eqref{eq:examp1} is the well-known ``water-filling" policy, where the water level satisfying \eqref{KKT_nu} and \eqref{eq:stat} does not have closed-form expression and has to be obtained from binary search in \cite{WirelessCom}. On the other hand, if the closed-form expression of $\bm{x}(\bm{\theta})$ cannot be obtained, one has to employ the FEM with extremely high complexity for finding the numerical result of the integration in \eqref{KKT_nu}.

In what follows, we resort to unsupervised deep learning to solve problem \eqref{prob:GeneralFunc}.

\subsubsection{Learning Approach}
To deal with the constraints of a problem, one can solve its primal-dual problem.
In particular, if problem \eqref{prob:GeneralFunc} is convex and the Slater's condition holds, then it is equivalent to the following problem \cite{boyd, gregory2018constrained},
\begin{align}   \label{prob:FuncOptLag}
    &\mathop \mathrm{max} \limits_{v_i(\bm{\theta}), {\lambda}_j} \mathop \mathrm{min} \limits_{\bm{x}(\bm{\theta})} \;  L \\
    &~~\text{s.t.}~~ \eqref{con:lambda} \nonumber
\end{align}

The Slater's condition generally holds in optimization problems with continues variables. However, the convexity does not hold in many cases.
If the problem is non-convex, a local optimal solution of problem \eqref{prob:GeneralFunc} can be obtained by solving problem \eqref{prob:FuncOptLag} \cite{luenberger1997optimization}.

To find the solution with unsupervised deep learning, we approximate the two functions in $L$, $\bm{x}(\bm{\theta})$ and $\bm{{v}}(\bm{\theta}) \triangleq [{{v}}_1(\bm{\theta}),...,{{v}}_I(\bm{\theta})]^{\rm T}$, by two DNNs denoted as $\mathcal{N}_x (\bm{\theta};\bm{\omega}_x)$ and $\mathcal{N}_{v} (\bm{\theta};\bm{\omega}_{v})$ respectively with  model parameters $\bm{\omega}_x$ and $\bm{\omega}_{v}$. According to the Universal Approximation Theory, a deterministic continuous function defined over a compact set can be approximated by a DNN, and the approximation can be arbitrarily accurate \cite{Hornik1989UnivApprox}.
By replacing $\bm{x}(\bm{\theta})$ and $\bm{{v}}(\bm{\theta})$ with $\hat{\bm{x}}(\bm{\theta}) \!\triangleq\! \mathcal{N}_x (\bm{\theta};\bm{\omega}_x)$ and $\hat{\bm{{v}}}(\bm{\theta}) \!\triangleq\! \mathcal{N}_{v} (\bm{\theta};\bm{\omega}_{v})$, problem \eqref{prob:FuncOptLag} can be re-written as,
\begin{align}   \label{prob:DNN}
    \mathop \mathrm{max} \limits_{{\bm{\omega}}_{v}, {\lambda}_j} \mathop \mathrm{min} \limits_{\bm{\omega}_x} \quad & \hat{L} = {\mathbb{E}}_{\bm{\theta}} \left\{ f\left[\bm{\hat{x}}(\bm{\theta}),\bm{\theta}\right] +
    \sum\limits_{i = 1}^I {\hat{v}_i(\bm{\theta}) C_i\left[\bm{\hat{x}}(\bm{\theta}),\bm{\theta}\right]} +
    \sum_{j=I+1}^{I+J}{{\lambda}_j{\left\{ C_j\left[\bm{\hat{x}}(\bm{\theta}),\bm{\theta}\right] \right\}}} \right\}\\
    \text{s.t.} \quad & \hat{{v}}_i(\bm{\theta}) \geq {0},  \forall {\bm{\theta} \in \mathcal{D}_{\theta}}, {\lambda}_j \geq 0. \label{con:hlambda}\tag{\theequation a}
\end{align}

Then, the primal-dual method can be used to update primal variables $\bm{\omega}_x$, dual variables ${\bm{\omega}}_{v}$, and the Lagrange multiplier for statistical constraint ${\lambda}_j$ iteratively
to find a solution of problem \eqref{prob:DNN}.
At the $t$th iteration, these variables can be updated by the stochastic gradient descent (SGD) method and the stochastic gradient ascent (SGA) method as
\begin{align}
\bm{\omega}_x^{(t+1)} &= \bm{\omega}_x^{(t)} -\phi_{\omega_{x}}(t){\nabla_{{\omega_{x}}}}{{\hat L}^{(t)}}, \label{eq:DLx}\\
{\bm{\omega}}_{v}^{(t+1)} &={\bm{\omega}}_{v}^{(t)}+ \phi_{\omega_{{v}}}(t) {\nabla_{{\bm{\omega}}_{v}}}{{\hat L}^{(t)}}, \label{eq:DLlamda} \\
{{{\lambda}}}_j^{(t+1)} &= \left( {{{\lambda}}}_j^{(t)} + \phi_{{{\lambda}}_j}(t) \frac{\partial{\hat L}^{(t)}}{\partial{{{\lambda}}_j}} \right)^+, \label{eq:DLnu}
\end{align}
where ${\left(x\right)}^+ \!\triangleq\! \max\!{\left\{x,0\right\}}$ ensures ${{{\lambda}}}_j^{(t+1)}>0$, $\phi_{\omega_{x}}(t)$, $\phi_{\omega_{{v}}}(t)$ and $\phi_{{{\lambda}}_j}(t)$ are the learning rates for updating $\bm{\omega}_x$, ${\bm{\omega}}_{v}$ and ${\lambda}_j$, and ${\nabla _{\omega_x}}  \hat{L}^{(t)}$ and ${\nabla _{\omega_{v}}} \hat{L}{^(t)}$ are the gradients of $\hat{L}$ w.r.t. ${\omega_x}$ and ${\omega_{v}}$, respectively.\footnote{The gradient of a scalar $x$ w.r.t. to a vector ${\bm{y}}_{N_y \times 1}$ is defined as $\nabla_{{\bm{y}}}x \triangleq [{\partial{x}}/{\partial{y_1}},...,{\partial x}/{\partial y_{N_y}}]^{\rm T}$ and the gradient of a vector $\bm{x}_{N_x \times 1}$ w.r.t. to a vector ${\bm{y}}_{N_y \times 1}$ is defined as $\nabla_{{\bm{y}}} \bm{x} \triangleq [\nabla_{{\bm{y}}} {x_1},...,\nabla_{{\bm{y}}} {x_{N_x}}]$.}
The method to compute the gradient and derivative is provided in Appendix \ref{App:derive}.

To guarantee $\hat{{v}}_i(\bm{\theta}) \geq {0}, \forall {\bm{\theta} \in \mathcal{D}_{\theta}}$, we need to choose a proper activation function in the output layer of $\mathcal{N}_{v} (\bm{\theta};\bm{\omega}_{v})$, e.g., $\texttt{ReLU}(x) \triangleq \max(x,0)$ or $\texttt{SoftPlus}(x) \triangleq \ln[1+\exp(x)]$ \cite{nair2010rectified,glorot2011deep}.

The DNNs are trained by optimizing $\bm{\omega}_x$, $\bm{\omega}_{v}$, ${\lambda}_j$, $j=I+1,...,I+J$ respectively with the SGD and SGA methods. As shown in \cite{eisen2019learning}, the primal-dual method converges at least to a local optimal solution of the primal-dual problem of the original problem. A local optimal solution is either at a stationary point of $\hat{L}$ or on the boundary of the feasible region. Thus, the following properties hold for the obtained solutions: $\nabla_{\bm{\omega}_x} \hat{L}= \bm{0}$, $\nabla_{\bm{\omega}_{v}} \hat{L} = \bm{0}$ (or $\hat{{v}_i}(\bm{\theta}) = 0$) and ${\partial\hat{L}}/{\partial{\lambda}_j} =0$ (or ${\lambda}_j=0$), $j=I+1,...,I+J$. These properties implicitly serve as the ``supervised signal" of the DNNs.
Since the DNNs are trained without labels, it belongs to unsupervised learning.


\section{Resource Allocation with Unsupervised Learning in URLLC} \label{sec:system_model}
In this section, we illustrate how to apply the framework presented in previous section. To this end, we minimize the bandwidth required by satisfying the QoS of every user with URLLC by optimizing bandwidth allocation with or without dynamic power allocation.

For the policy without power allocation, the BS only allocates bandwidth among users according to their large-scale channel gains, which is formulated as a variable optimization problem. For the policy with power allocation, the BS also adjusts transmit power according to the small-scale channel gains of the users, which is formulated as a hybrid variable and functional optimization problem, where the ``variables" are in two timescales.

\subsection{System Model and QoS Constraints}
\subsubsection{System, Traffic and Channel models}
Consider a DL orthogonal frequency division multiple access system, where a BS with $N_\mathrm{t}$ antennas serves $K$ single-antenna users. The maximal transmit power and the total bandwidth of the BS are denoted by $P_\mathrm{max}$ and $W_{\max}$, respectively.

The packets for each user arrive at the buffer of the BS randomly. The inter-arrival time between packets could be shorter than the service time of each packet. Therefore, the packets may wait in the buffer of the BS. We consider a queueing model that the packets for different users wait in different queues and are served according to a first-come-first-serve order.

Time is discretized into slots, each with duration $T_\mathrm{s}$. The duration for DL data transmission in one time slot is $\tau <T_\mathrm{s}$. Since the E2E delay requirement in URLLC is typically shorter than the coherence time of small-scale channel, the channel is quasi-static and time diversity cannot be exploited. To improve the transmission reliability within the delay bound, we consider frequency hopping, where each user is assigned with different subchannels in adjacent slots. When the frequency interval between adjacent subchannels is larger than the coherence bandwidth, the small-scale channel gains of a user among slots are mutual independent.
Since the packet size $u$ in URLLC is typically small (e.g., $20$~bytes or $32$~bytes \cite{3GPP2016Scenarios}), the bandwidth required for transmitting each packet is less than the channel coherence bandwidth. Therefore, the small-scale channel is flat fading.

\begin{figure}[ht]
        \centering
        \begin{minipage}[t]{0.5\textwidth}
        \includegraphics[width=1\textwidth]{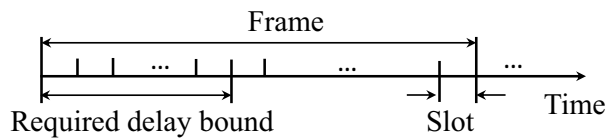}
        \end{minipage}
        \caption{Large-scale channels change among frames, and small-scale channels change among time slots due to frequency hopping.}
        \label{fig:TwoTimescale}
        \vspace{-0.2cm}
\end{figure}

As shown in \cite{WirelessCom}, the large-scale channel gain of a user varies when its moving distance is comparable to the decorrelation distance of shadowing, i.e., $50\sim100$~m. Thus, the coherence time of the large-scale channel gain is around a few seconds, much longer than the delay bound $D_\mathrm{max}$ and the slot duration $T_\mathrm{s}$ (e.g., in 5G New Radio, $T_\mathrm{s}$ can be much shorter than $1$~ms \cite{3GPP2017Agree}). We assume that large-scale channel gains stay constant in each frame that consists of $N_{\rm f}$ time slots, and may vary in different frames. The relations among the timescales of the frames, slots, and the required delay bound are illustrated in Fig. \ref{fig:TwoTimescale}.

In URLLC, the blocklength of channel coding is short due to the short transmission duration, and hence the impact of decoding errors on reliability cannot be ignored. Since Shannon's capacity formula cannot be employed to characterize the probability of decoding errors \cite{Gross2015Delay}, we consider the achievable rate in finite blocklength regime. In quasi-static flat fading channels, when small-scale channel gain is available at the transmitter and receiver, the achievable rate of the $k$th user can be accurately approximated by \cite{Yury2014Quasi},
\begin{align}   \label{eq:Srv}
    s_k \approx \frac{\tau W_k}{u \ln{2}} \left[\ln\left(1+\frac{\alpha_k g_k P_k}{ N_0 W_k}\right) - \sqrt{\frac{V_k}{\tau W_k}} Q_\mathrm{G}^{-1}\!\left({\varepsilon^\mathrm{c}_k}\right)\right] \ \text{( packets/slot)},
\end{align}
where $W_k$ and $P_k$ are the bandwidth and the transmit power allocated to the $k$th user, respectively, $\varepsilon^\mathrm{c}_k$ is the decoding error probability of the $k$th user, $\alpha_k$ and $g_k$ are the large-scale channel gain and small-scale channel gain of the $k$th user, respectively, $Q_\mathrm{G}^{-1} (x)$ is the inverse of the Gaussian Q-function, and $V_k$ is the channel dispersion given by $V_k=1-\frac{1}{\left[1+\frac{\alpha_k g_k P_k}{N_0 W_k }\right]^2}$ \cite{Yury2014Quasi}.

Although the achievable rate is in closed-form, it is still too complicated to obtain graceful results. As shown in \cite{Gross2015Delay}, if the SNR $\frac{\alpha_k g_k P_k}{N_0 W_k } \ge$ $5$~dB, $V_k \!\approx\! 1$ is accurate. Since high SNR is required to ensure ultra-high reliability and ultra-low latency, such approximation is reasonable. Even when the SNR is not high, we can obtain a lower bound of the achievable rate by substituting $V_k \!\approx\! 1$ into $s_k$. Then, the required $\varepsilon^\mathrm{c}$ can be satisfied if the lower bound of \eqref{eq:Srv} is used to characterize the achievable rate.

\subsubsection{Reliability and Delay Constraints}


The QoS requirement of each user  can be characterized by an E2E delay bound $D_\mathrm{max}$ for each packet and the overall packet loss probability $\varepsilon_\mathrm{max}$.

The uplink transmission delay, backhaul delay and processing delay have been studied in \cite{She2018Joint}, \cite{Gongzheng2016Backhaul} and \cite{Makki2018FastHARQ}, respectively, and can be subtracted from the E2E delay. In this paper, $D_\mathrm{max}$ is the DL delay, which consists of the queueing delay (denoted as $D^\mathrm{q}_k$ for the $k$th user), transmission delay $D^\mathrm{t}$ (which equals to $T_s$, including the data transmission time $\tau$ and the channel training time) and decoding delay $D^\mathrm{c}$.

$D^\mathrm{t}$ and $D^\mathrm{c}$ are constant values depending on the standardization and hardware \cite{Condoluci2017Reserv}. Due to the random packet arrival, $D^\mathrm{q}_k$ is random.
To ensure the delay requirement, $D^\mathrm{q}_k$ should be bounded by $D^\mathrm{q}_\mathrm{max} \!\triangleq\! D_\mathrm{max} \!-\! D^\mathrm{t} \!-\! D^\mathrm{c}$ with a very low probability, because a packet will be useless if the queueing delay of the packet exceeds $D^\mathrm{q}_\mathrm{max}$.

Denote $\varepsilon^\mathrm{q}_k \!\triangleq\! \Pr\{D^\mathrm{q}_k \!>\! D^\mathrm{q}_\mathrm{max}\}$ as the queueing delay violation probability. Then, the overall reliability requirement can be characterized by $1 - (1 - \varepsilon^\mathrm{c}_k) (1 - \varepsilon^\mathrm{q}_k) \approx \varepsilon^\mathrm{c}_k + \varepsilon^\mathrm{q}_k \leq \varepsilon_\mathrm{max}$. This approximation is very accurate, because the values of $\varepsilon^\mathrm{c}$ and $\varepsilon^\mathrm{q}$ are very small in URLLC.

Effective bandwidth and effective capacity have been widely used to analyze the tail probability of queueing delay, i.e., $D^\mathrm{q}_\mathrm{max}$ is large or $\varepsilon^\mathrm{q}_k$ is extremely small \cite{EB,EC}.
As analyzed in \cite{She2018CrossLayer}, if the slot duration is much shorter than the delay bound, which is true in URLLC, effective bandwidth can be used to analyze the queueing delay at the BS for Poisson, interrupted Poisson and switched Poisson arrival processes.

We take the Poisson arrival process with the average packet arrival rate $a_k$~packets/slot as an example, whose effective bandwidth can be expressed as \cite{She2018CrossLayer}
\begin{align}   \label{eq:EB}
    B^\mathrm{E}_k = \frac{\ln{(\varepsilon_\mathrm{max}/2)}}{D^\mathrm{q}_\mathrm{max}\ln \!{\left[1-\frac{\ln{(\varepsilon_\mathrm{max}/2)}}{a_k D^\mathrm{q}_\mathrm{max}}\right]}} \; \text{(packets/slot)}.
\end{align}
If the constant packet service rate (i.e., the achievable rate) of the $k$th user is no less than $B^\mathrm{E}_k$, then we have $\Pr\{D^{\rm q}_k \!>\! D^{\rm q}_{\max}\} \!\leq\! \exp\{-\vartheta_k B^\mathrm{E}_k D^\mathrm{q}_\mathrm{max} \}$, where $\vartheta_k$ is the QoS component, which reflects the decay rate of the tail probability of the queueing delay. By setting the upper bound in the inequity equals to $\varepsilon_\mathrm{max}/2$, we can obtain
\begin{align}
\vartheta_k = \ln \!{\left[1-\frac{\ln{(\varepsilon_\mathrm{max}/2)}}{a_k D^\mathrm{q}_\mathrm{max}}\right]}.\label{eq:vartheta}
\end{align}

Since the small-scale channel gains of a user are independent among slots owing to frequency hopping, the effective capacity of the $k$th user can be expressed as \cite{Tang2007TWC}
\begin{align}   \label{eq:EC}
    C^\mathrm{E}_k = - \frac{1}{\vartheta_k} \ln{\mathbb{E}_{{g_k}} \left\{e^{- \vartheta_k s_k}\right\}} \; \text{(packets/slot)}.
\end{align}

When both the packet arrival process and the packet service process are stochastic, $D^\mathrm{q}_\mathrm{max}$ and $\varepsilon^\mathrm{q}_k$ can be satisfied if \cite{Lingjia2007TIT}
\begin{align}
C^\mathrm{E}_k \!\geq\! B^\mathrm{E}_k.\label{eq:EBEC}
\end{align}

To simplify the optimization problem, we set $\varepsilon^\mathrm{c}_k \!=\! \varepsilon^\mathrm{q}_k \!=\! \varepsilon_\mathrm{max}/2$. The results in \cite{She2018CrossLayer,She2018Joint} show that the optimal values of $\varepsilon^\mathrm{c}_k $ and $\varepsilon^\mathrm{q}_k$ are in the same order of magnitude, {and the simplification will only lead to a negligible performance loss.} Then, the QoS of each user, characterized by $D_\mathrm{max}$ and $\varepsilon_\mathrm{max}$, can be satisfied if \eqref{eq:EBEC} holds after substituting the expression of $s_k$ in \eqref{eq:Srv} into \eqref{eq:EC}. Such a QoS constraint is complicated and may not be expressed in closed-form.

\subsection{Bandwidth Allocation: A Variable Optimization Problem}\label{BLOnly}
In this subsection, we assume that the transmit power does not change according to small-scale channel gains, which is reasonable in practical cellular networks where modulation and coding schemes are adjusted according to channel realizations with fixed power allocation \cite{3GPPDLpower}. We optimize the bandwidth allocation policy according to the large-scale channel gains of users. Hence, the environmental parameters can be expressed as ${\bm \theta} ={\bm \alpha} \!\triangleq\! [\alpha_1, \cdots\!, \alpha_K]^\mathrm{T}$. 

\subsubsection{Problem Formulation}
In particular, assume that $P_0 \!=\! P_{\max}/W_{\max}$. Then, by substituting $\varepsilon^\mathrm{c}_k \!=\! \varepsilon_\mathrm{max}/2$ into \eqref{eq:Srv}, the achievable rate of the $k$th user can be re-written as follows,
\begin{align}
s_k \!=\! \frac{\tau W_k}{u \ln{2}} \left[\ln\!\left(1\!+\!\frac{\alpha_k g_k }{ N_0 } P_0\right) \!-\!\frac{Q_\mathrm{G}^{-1}\!\left({\varepsilon_\mathrm{max}/2}\right)}{\sqrt{\tau W_k}}\right] \label{eq:Pconst}.
\end{align}

The bandwidth allocation problem can be formulated as a variable optimization problem that minimizes the total bandwidth required to ensure the QoS of every user, i.e.,
\begin{align}   \label{prob:RAconst}
    \mathop \mathrm{min} \limits_{W_k, k=1,...,K} \quad& \sum_{k=1}^{K} {W_k} \\
    \text{s.t.} \quad
    & \mathbb{E}_{g_k} \left\{e^{- \vartheta_k s_k}\right\} - e^{- \vartheta_k B^\mathrm{E}_k} \leq 0, k=1,...,K \label{con:QPconst} \tag{\theequation a} \\
    & \sum_{k=1}^{K} {W_k} \leq W_{\max}, \label{con:Wmax} \tag{\theequation b} \\
    & W_k \geq 0, k=1,...,K, \nonumber
\end{align}
where \eqref{con:QPconst} is obtained by substituting \eqref{eq:EC} into \eqref{eq:EBEC},
$s_k$ is given by \eqref{eq:Pconst},
and $W_{\max}$ is the maximal total bandwidth.

Since the left-hand side of the constraint in \eqref{con:Wmax} is the same as the objective function, we can remove it when solving problem \eqref{prob:RAconst}. If the minimal bandwidth required to guarantee the QoS requirement of every user exceeds $W_{\max}$, problem \eqref{prob:RAconst} will be infeasible. After removing the constraint in \eqref{con:Wmax}, the bandwidth allocation of every user is mutually independent among each other. Thus, problem \eqref{prob:RAconst} can be equivalently decomposed into $K$ single-user problems,
\begin{align}
\mathop \mathrm{min} \limits_{W_k} \quad& {W_k} \label{eq:SUW}\\
    \text{s.t.} \quad
    & \eqref{con:QPconst}, W_k \geq 0. \nonumber
\end{align}
In the rest part of this subsection, the index $k$ is omitted for notational simplicity.

\subsubsection{Optimizing $W$ from the Variable Optimization Problem}
To provide a baseline for the unsupervised deep learning method, we first find the optimal solution of problem in \eqref{eq:SUW} for any given realizations of the environmental parameters.

Since $s_k$ in \eqref{eq:Pconst} increases with $W$, the left-hand side of \eqref{con:QPconst} decreases with $W$, and the minimal bandwidth is obtained when the equality in \eqref{con:QPconst} holds. If effective capacity can be derived as a closed-form expression, say in large-scale antenna systems \cite{ECTCOM15}, then we can use binary search to find the minimal bandwidth. In general wireless systems, the effective capacity does not have closed-form expression, and hence \eqref{con:QPconst} cannot be expressed in closed form. To find the optimal bandwidth allocated to each user, one can use stochastic optimization through the following iterations,
\begin{align}   \label{opt:W}
    W^{(t+1)} = {\left[W^{(t)} + \phi(t) \left(e^{- \vartheta s^{(t)}} - e^{-\vartheta B^\mathrm{E}}\right)\right]}^+,
\end{align}
where $\phi(t) \!>\! 0$ is the learning rate, $s^{(t)}$ is the achievable rate computed from \eqref{eq:Pconst} given the realization of $g$ in the $t$th iteration, and one realization of $g$ can be obtained in each slot.
With $\phi(t) \!\sim\! \mathcal{O}\!\left(\frac{1}{t}\right)$, $\{W^{(t)}\}$ converges to the unique optimal bandwidth \cite{Bottou1998Stochastic} thanks to the monotonicity of the function of the left-hand side of \eqref{con:QPconst}.

\subsubsection{Optimizing $W(\alpha)$ with Unsupervised Deep Learning}\label{Unsurp-1}
For the sake of learning to optimize problem \eqref{eq:SUW}, we first formulate a functional optimization problem of finding the mapping from $\alpha$ to the optimal solution of problem \eqref{eq:SUW} as follows,
\begin{align}   \label{prob:RA_FuncOpt}
    \mathop \mathrm{min} \limits_{W(\alpha)} \quad& \mathbb{E}_{\alpha} \left\{W(\alpha)\right\} \\
    \text{s.t.} \quad & \mathbb{E}_{g} \left\{e^{- \vartheta s[W(\alpha);\alpha]}\right\} - e^{- \vartheta B^\mathrm{E}} \leq 0, \label{eq:EBECalpha}\tag{\theequation a} \\
    &W(\alpha) \geq 0, \nonumber
\end{align}
where $s[W(\alpha);\alpha] \!=\! \frac{\tau W(\alpha)}{u \ln{2}} \left[\ln\!\left(1\!+\!\frac{\alpha g }{ N_0 } P_0\right) \!-\!\frac{Q_\mathrm{G}^{-1}\!\left({\varepsilon_\mathrm{max}/2}\right)}{\sqrt{\tau W(\alpha)}}\right]$ is the re-written expression of \eqref{eq:Pconst}, and \eqref{eq:EBECalpha} is an instantaneous constraint although it consists of expectation, because the expectation is taken over small-scale channel gains for a given realization of the environment parameter ${\alpha}$.
The constraint in \eqref{eq:EBECalpha} is non-convex, hence problem \eqref{prob:RA_FuncOpt} is non-convex. According to the discussion in Section II-D, a local optimal solution of problem \eqref{prob:RA_FuncOpt} can be found by solving its primal-dual problem,
\begin{align}
    \mathop \mathrm{max} \limits_{{v}(\alpha)} \mathop \mathrm{min} \limits_{W(\alpha)} \ & L_1 \!\triangleq\! \mathbb{E}_{\alpha} \!\left\{W(\alpha) \!+\! {v}(\alpha) \!\left( \mathbb{E}_{{g}} \!\left\{\!e^{- \vartheta s[W(\alpha);\alpha]}\!\right\} \!-\! e^{-\vartheta B^\mathrm{E}} \right)\right\} \label{eq:PD} \\
    \text{s.t.} \ &  W(\alpha) {\geq} 0, \  {v}(\alpha) > 0, \forall \alpha > 0, \nonumber
\end{align}
where ${v}(\alpha)$ is the Lagrange multiplier function. The constraint $W(\alpha) \!\geq\! 0$ and the corresponding Lagrange multiplier are not included in $L_1$, because the optimal bandwidth is always positive and the corresponding Lagrange multiplier is always zero.

To apply the framework in Section \ref{subsec:funcDNN} to solve problem \eqref{eq:PD}, we approximate the functions $W(\alpha)$ and ${v}(\alpha)$ by two DNNs, denoted as $\hat{W} \!\triangleq\! \mathcal{N}_{W} \left(\alpha;\bm{\omega}_{W}\right)$ and $\hat{{v}} \!\triangleq\! \mathcal{N}_{{v}} \left(\alpha;\bm{\omega}_{{v}}\right)$, respectively.
By using appropriate activation function in the output layers of both DNNs, $\hat{W}$ and $\hat{{v}}$ are positive. The model parameters of the DNNs, $\bm{\omega}_{W}$ and $\bm{\omega}_{{v}}$, can be obtained iteratively as follows,
\begin{align}
    \bm{\omega}_W^{(t+1)} \!&=\! \bm{\omega}_W^{(t)} \!-\! \phi_{\bm{\omega}_W}(t) \nabla_{\bm{\omega}_W} \hat{L}_1^{(t)} = \bm{\omega}_W^{(t)} \!-\! \frac{\phi_{\bm{\omega}_W}(t)}{N_\mathrm{b}} \sum_{n=1}^{N_\mathrm{b}} {\left[ \nabla_{\bm{\omega}_W} \mathcal{N}_W\!\left(\alpha^{(t,n)};\bm{\omega}_W^{(t)}\right) \frac{\mathrm{d} \hat{L}_1^{(t)}}{\mathrm{d} \hat{W}^{(t,n)}} \right]}, \label{trn:ParaW} \\
    \bm{\omega}_v^{(t+1)} \!&=\! \bm{\omega}_v^{(t)} \!+\! \phi_{\bm{\omega}_v}(t) \nabla_{\bm{\omega}_{v}} \hat{L}_1^{(t)} = \bm{\omega}_v^{(t)} \!+\! \frac{\phi_{\bm{\omega}_v}(t)}{N_\mathrm{b}} \sum_{n=1}^{N_\mathrm{b}} {\left[ \nabla_{\bm{\omega}_v} \mathcal{N}_v \!\left(\alpha^{(t,n)};\bm{\omega}_W^{(t)}\right) \frac{\mathrm{d} \hat{L}_1^{(t)}}{\mathrm{d} \hat{v}^{(t,n)}} \right]}, \label{trn:ParaLambda}
\end{align}
where $\hat{L}_1^{(t)} \!\triangleq\! \frac{1}{N_\mathrm{b}} \sum_{n=1}^{N_\mathrm{b}} {\left[\hat{W}^{(t,n)} \!+\! \hat{v}^{(t,n)} \!\left( e^{-\! \vartheta \hat{s}^{(t,n)}} \!\!-\! e^{-\vartheta B^\mathrm{E}} \right)\right]}$ is the estimated objective function in \eqref{eq:PD} with $N_\mathrm{b}$ realizations of large-scale channel gains while $\alpha^{(t,n)}$ and $\hat{s}^{(t,n)}$ are respectively the $n$th realizations of the large-scale channel gain and the achievable rate in the $t$th iteration, $\hat{W}^{(t,n)} \!\triangleq\! \mathcal{N}_{W} \left(\alpha^{(t,n)};\bm{\omega}^{(t)}_{W}\right)$ and $\hat{v}^{(t,n)} \!\triangleq\! \mathcal{N}_{v} \left(\alpha^{(t,n)};\bm{\omega}^{(t)}_{v}\right)$.
In \eqref{trn:ParaW} and \eqref{trn:ParaW}, the derivative of $\hat{L}_1^{(t)}$ w.r.t. $\hat{W}^{(t,n)}$ and $\hat{v}^{(t,n)}$ can be derived as follows,
\begin{align}
\frac{\mathrm{d} \hat{L}_1^{(t)}}{\mathrm{d} \hat{W}^{(t,n)}} = 1 - \hat{{v}}^{(t,n)} \vartheta \frac{\partial \hat{s}^{(t,n)}}{\partial \hat{W}^{(t,n)}} e^{- \vartheta \hat{s}^{(t,n)}}, ~~
\frac{\mathrm{d} \hat{L}_1^{(t)}}{\mathrm{d} \hat{v}^{(t,n)}} = e^{-\! \vartheta \hat{s}^{(t,n)}} \!\!-\! e^{-\vartheta B^\mathrm{E}},\nonumber
\end{align}
where the values of $B^{\rm E}$ and $\vartheta$ are computed according to \eqref{eq:EB} and \eqref{eq:vartheta}, respectively, and
\begin{align}
\frac{\partial \hat{s}^{(t,n)}}{\partial \hat{W}^{(t,n)}} = \frac{1}{{u\ln 2}}\left[ {\tau \ln \left( {1 + \frac{{{\alpha ^{\left( {t,n} \right)}}g{P_0}}}{{{N_0}}}} \right) - \frac{{Q_{\rm{G}}^{ - 1}\left( {{\varepsilon _{{\rm{max}}}}/2} \right)}}{2}\sqrt {\frac{\tau }{\hat{W}^{(t,n)}}} } \right].\nonumber
\end{align}
The gradient matrices $\nabla_{\bm{\omega}_W} \mathcal{N}_W\left(\alpha^{(t,n)};\bm{\omega}_W^{(t)}\right)$ and $\nabla_{\bm{\omega}_{v}} \mathcal{N}_{v}\left(\alpha^{(t,n)};\bm{\omega}_{v}^{(t)}\right)$ can be computed
by backward propagation.

After the iterations converge, we can obtain a well-trained DNN $\mathcal{N}_{W} \left(\alpha;\bm{\omega}_{W}\right)$, which can approximate the optimal function of $W(\alpha)$. Then, the BS only needs to compute the bandwidth allocated to each user from $\mathcal{N}_{W} \left(\alpha;\bm{\omega}_{W}\right)$ after obtaining the large-scale channel gain of each user at the beginning of each frame.

\subsection{Bandwidth and Power Allocation: A Hybrid Variable and Functional Optimization Problem}
In this subsection, we illustrate how to solve a functional optimization problem subject to both instantaneous and statistic constraints. Although the BSs in the fourth generation cellular systems do not adjust transmit power according to small-scale channel, the total bandwidth required by URLLC can be further reduced with dynamic power allocation. We optimize bandwidth allocation according to the large-scale channel gains of  multiple users (i.e., ${\bm \theta} ={\bm \alpha}$) and power allocation according to their small-scale channel gains (i.e., ${\bm \theta} ={\bm g}) \triangleq [g_1,\cdots\!, g_K]^\mathrm{T}$). Hence, the jointly optimized policy operates in two timescales.

\subsubsection{Problem Formulation}

To reflect the impact of the two-timescale resource allocation, we re-write the achievable rate of the $k$th user in \eqref{eq:Srv} to satisfy $\varepsilon^\mathrm{c}_k = \varepsilon_\mathrm{max}/2$ as,
\begin{align}
s_k [W_k, P_k(\bm{g});g_k] \!=\! \frac{\tau W_k}{u \ln{2}} \left[\ln\!\left(1 \!+\! \frac{\alpha_k g_k P_k(\bm{g})}{ N_0 W_k}\right) \!-\! \frac{Q_\mathrm{G}^{-1}\!\left({\varepsilon_{\max}/2}\right)}{\sqrt{\tau W_k}}\right]. \label{con:Srv}
\end{align}

The problem of joint bandwidth and power allocation that minimizes the total bandwidth required to ensure the QoS under the constraint of maximal power $P_\mathrm{max}$ can be formulated as,
\begin{align}   \label{prob:RA}
    \mathop \mathrm{min} \limits_{W_k, P_k(\bm{g})} \quad& \sum_{k=1}^{K} {W_k} \\
    \text{s.t.} \quad
    & \mathbb{E}_{\bm{g}} \left\{e^{- \vartheta_k s_k[W_k, P_k(\bm{g});g_k]}\right\} - e^{-\vartheta_k B^\mathrm{E}_k} \leq 0, ~k=1,\cdots, K\label{con:Q} \tag{\theequation a} \\
    & \sum_{k=1}^{K} {P_k(\bm{g})} \leq P_\mathrm{max}, \label{con:Pmax} \tag{\theequation b}\\
    & W_k \geq 0, P_k(\bm{g}) \geq 0, ~k=1,\cdots, K.  \label{con:positive} \tag{\theequation c}
\end{align}
The left-hand side of \eqref{con:Q} is a function of $P_k(\bm{g})$, which measures the QoS requirement in each frame and depends on the distribution of environmental parameters $\bm{g}$. Thus, \eqref{con:Q} are the statistic constraints for the functional optimization. The constraints in \eqref{con:Pmax} and \eqref{con:positive} only depend on specific realizations of environmental parameters, and hence are instantaneous constraints. The total bandwidth constraint is removed as explained in previous subsection. If the minimal total bandwidth is higher than $W_{\max}$, then the problem is infeasible.

This a generic functional optimization problem, including both functional optimization for $P_k(\bm{g})$ and variable optimization for $W_k$. In what follows, we apply the proposed framework to solve this hybrid variable and functional optimization problem.




\subsubsection{Optimizing $W_k$ and $P_k({\bf g})$ from Necessary Conditions} \label{sec:Sym}
To provide a baseline for the learning-based solution, we first derive the optimal solution of problem \eqref{prob:RA} from the necessary conditions. To simplify the notation, in the sequel we again use $s_k$ to denote $s_k[W_k, P_k(\bm{g});g_k]$ in \eqref{con:Srv}.

The Lagrangian of problem \eqref{prob:RA} can be expressed as follows,
\begin{align}
L_2 &\! \triangleq\! \sum_{k=1}^{K} {W_k} \!+\! \sum_{k=1}^{K} {\lambda_k \!\left( \mathbb{E}_{\bm{g}} \!\left\{\!e^{- \vartheta_k s_k}\!\right\} \!-\! e^{-\vartheta_k B^\mathrm{E}_k} \right)}+ \!\!\int_{\mathbb{R}_+^K} \!\!\!{\left[h(\bm{g}) \!\left(\sum_{k=1}^{K} {P_k(\bm{g})} \!-\! P_\mathrm{max}\!\right) - \sum_{k=1}^{K} {v_k(\bm{g}) P_k(\bm{g})}\right] \!\mathrm{d}\bm{g}} \nonumber
\end{align}
where $\lambda_k$, $h(\bm{g})$ and $v_k(\bm{g})$ are the Lagrange multipliers. Similar to the Lagrangian in \eqref{eq:PD}, the constraint $W_k \!\geq\! 0$ and the corresponding Lagrange multiplier are omitted in $L_2$.

Then, the optimal solution of problem \eqref{prob:RA} should satisfy its {first-order necessary conditions}, which can be derived as \cite{gregory2018constrained},
\begin{align}
    &\frac{\mathrm{\partial} L_2}{\mathrm{\partial} P_k(\bm{g})} =  h(\bm{g}) - v_k(\bm{g}) - \lambda_k \vartheta_k \frac{\partial s_k}{\partial P_k(\bm{g})} e^{- \vartheta s_k} p(\bm{g}) = 0, \label{cnd:P} \\
    &\frac{\partial L_2}{\partial W_k} = 1 - \lambda_k \vartheta_k \mathbb{E}_{\bm{g}} \left\{\frac{\partial s_k}{\partial W_k} e^{- \vartheta s_k}\right\} = 0, \label{cnd:W} \\
    &\lambda_k \!\left( \mathbb{E}_{\bm{g}} \!\left\{\!e^{- \vartheta_k s_k}\!\right\} \!-\! e^{-\vartheta_k B^\mathrm{E}_k} \right) = 0,\label{cnd:kappa}\\
    & h(\bm{g}) \!\left(\sum_{k=1}^{K} {P_k(\bm{g})} \!-\! P_\mathrm{max}\!\right) = 0, \forall \bm{g} \in {\mathbb{R}_+^K}, \label{cnd:hg}\\
    & v_k(\bm{g}) P_k(\bm{g}) = 0, \forall \bm{g} \in {\mathbb{R}_+^K}, \label{cnd:vg}\\
    & W_k \geq 0, \lambda_k \geq 0, P_k(\bm{g}) \geq 0, h(\bm{g}) \geq 0, {v_k(\bm{g}) \geq 0}, \forall \bm{g} \in {\mathbb{R}_+^K},\label{cnd:positive}\\
    &\eqref{con:Q}\;\text{and}\; \eqref{con:Pmax}, \nonumber
\end{align}
where $p(\bm{g})$ is the joint PDF of $\bm{g}$.

{\bf Optimal Power Allocation:}
From \eqref{cnd:P} and \eqref{con:Srv} we have
\begin{align}   \label{eq:h}
    h(\bm{g}) &\!=\! \lambda_{k} \vartheta_{k} \frac{\partial s_k}{\partial P_k(\bm{g})} e^{- \vartheta s_k} p(\bm{g}) + v_k(\bm{g}) \nonumber \\
    &\!=\! \lambda_{k} \vartheta_{k} \frac{\tau W_{k}}{u \ln{2}} \frac{\alpha_{k} g_k}{N_0 W_{k}} \frac{1}{(1 \!+\! \gamma_k)} e^{- \vartheta s_k} p(\bm{g}) + v_k(\bm{g}) \nonumber \\
    &\!=\! \frac{\lambda_{k} \vartheta_{k} \alpha_{k} g_k \tau}{N_0 u \ln\!{2} \left(1 \!+\! \gamma_k\right)} {\left(1 \!+\! \gamma_k\right)}^{-\frac{\vartheta_{k} \tau W_{k}}{u \ln\!{2}}} e^{\frac{\vartheta_{k} \sqrt{\tau W_{k}} Q_\mathrm{G}^{-1}\!\left({\varepsilon_\mathrm{max}/2}\right)}{u \ln\!{2}}} p(\bm{g}) + v_k(\bm{g}) \nonumber \\
    &\!=\! \frac{\beta_{k} g_k p(\bm{g})}{{\left(1 \!+\! \gamma_k\right)}^{\frac{1}{\eta_{k}}}} + v_k(\bm{g}),
\end{align}
where $\gamma_k \! \triangleq \! \frac{\alpha_{k} g_k P_k(\bm{g})}{N_0 W_{k}}$ is the SNR of the $k$th user, $\beta_{k} \! \triangleq \! \frac{\lambda_{k} \vartheta_{k} \alpha_{k} \tau}{N_0 u \ln\!{2}} e^{\frac{\vartheta_{k} \sqrt{\tau W_{k}} Q_\mathrm{G}^{-1}\!\left({\varepsilon_\mathrm{max}/2}\right)}{u \ln\!{2}}}$, and $\eta_{k} \! \triangleq \! 1/{\left({1+\frac{\vartheta_{k} \tau W_{k}}{u \ln\!{2}}}\right)}$.


From \eqref{eq:h}, we can see that if $h(\bm{g}) \!>\! \beta_{k} g_k p(\bm{g}) \!\geq\! \beta_{k} g_k p(\bm{g})/(1+\gamma_k)^{1/\eta_{k}}$, then $v_k(\bm{g}) \!>\! 0$. To satisfy the condition in \eqref{cnd:vg}, we have $P_k(\bm{g}) \!=\! 0$. In the case that $h(\bm{g}) \!<\! \beta_{k} g_k p(\bm{g})$, if $P_k(\bm{g}) \!=\! 0$, then $v_k(\bm{g})$ will be negative, which contradicts with the constraint $v_k(\bm{g}) \!\geq\! 0$ in \eqref{cnd:positive}. Thus, we have $P_k(\bm{g}) \!>\! 0$. To meet the constraint in \eqref{cnd:vg}, $v_k(\bm{g}) \!=\! 0$. {It is not hard to see that when $h(\bm{g}) \!=\! \beta_k g_k p(\bm{g})$, the solution is $P_k(\bm{g}) \!=\! 0$ and $v_k(\bm{g}) \!=\! 0$. {Given the solutions in the above cases}, the optimal power allocation policy can be expressed as,}
\begin{align}   \label{eq:P}
    P_k(\bm{g}) = \frac{N_0 W_{k}}{\alpha_{k} g_k} \left[{\left( \frac{g_k}{g_{k}^\mathrm{th}(\bm{g})} \right)}^{\eta_{k}} - 1\right]^+,
\end{align}
where $g_{k}^{\rm th}(\bm{g})\triangleq \frac{h(\bm{g})}{\beta_{k} p(\bm{g})}$. Since {the bandwidth required to guarantee the QoS of each user decreases with the transmit power allocated to the user}, the optimal solution of problem \eqref{prob:RA} is obtained when the equality in \eqref{con:Pmax} holds.
Substituting \eqref{eq:P} into $\sum_{k=1}^{K} {P_k(\bm{g})}=P_\mathrm{max}$, we have
\begin{align}   \label{opt:h}
    \sum_{k=1}^{K} {\frac{N_0 W_{k}}{\alpha_{k} g_k} \left[{\left( \frac{g_k}{g_{k}^\mathrm{th}(\bm{g})} \right)}^{\eta_{k}} - 1\right]^+} = \sum_{k \in \mathbb{K}^+} {\frac{N_0 W_{k}}{\alpha_{k} g_k} \left[{\left( \frac{g_k}{g_{k}^\mathrm{th}(\bm{g})} \right)}^{\eta_{k}} - 1\right]} = P_\mathrm{max},
\end{align}
where $\mathbb{K}^+$ denotes the set of users with positive transmit power. Since the values of $\eta_k, k=1,\cdots\!,K$, differ among users, $g_k^\mathrm{th}(\bm{g})$ can not be obtained in a closed-form  expression.

\emph{A Symmetric Case:} When all users have identical large-scale channel gains (i.e., $\alpha_k \!=\! \alpha$) and have the same average packet arrival rate (i.e., $a_k \!=\! a$, and hence $\vartheta_k=\vartheta$),  $W_k$ and $\eta_k$ are identical for different users (i.e., $W_k=W$ and $\eta_k \!=\! \eta$). In this case, $g_k^\mathrm{th}(\bm{g}) \!=\! g^\mathrm{th}(\bm{g})$, which can be derived as follows,
\begin{align} \label{opt:h2}
    g^\mathrm{th}(\bm{g}) = \left(\frac{\frac{\alpha P_\mathrm{max}}{N_0 W} \!+\! \sum_{k \in \mathbb{K}^+} {{g_k}^{-1}}}{\sum_{k \in \mathbb{K}^+} {{g_k}^{\eta-1}}}\right)^{-\frac{1}{\eta}}.
\end{align}
Substituting \eqref{opt:h2} into \eqref{eq:P}, we can derive the optimal power allocation policy as
\begin{align}\label{opt:Pfinal}
P_k(\bm{g}) = \frac{N_0 W}{\alpha g_k} \left(\frac{\frac{\alpha g_k P_\mathrm{max}}{N_0 W} + g_k \sum_{k \in \mathbb{K}^+} {{g_i}^{-1}}}{{g_k}^{1-\eta} \sum_{k \in \mathbb{K}^+} {{g_i}^{\eta-1}}} - 1\right)^+.
\end{align}
It is worth noting that the {elements in $\mathbb{K}^+$ and the function $P_k(\bm{g})$ rely} on each other. To find the solution, for each given realization of $\bm{g}$ we compute \eqref{opt:Pfinal} and update $\mathbb{K}^+$ iteratively from the initial user set $\mathbb{K}^+_0 \!=\! \{1,2,\!\cdots,K\}$. According to the results of $P_k(\bm{g})$ obtained from \eqref{opt:Pfinal}, the users with negative transmit power are removed from $\mathbb{K}^+$. By repeating this procedure until $P_k(\bm{g}) \!\geq\! 0$ for all $k \!\in\! \mathbb{K}^+$, we can obtain the optimal power allocation policy.

{\bf Optimal Bandwidth Allocation:}
Due to the expectation in \eqref{con:Q} and the complex expression of $s_k$ in \eqref{con:Srv}, the optimal bandwidth allocation cannot be obtained in closed-form. The solution can be found with stochastic optimization using the iteration formula in \eqref{opt:W}, where $s_k^{(t)}$ is obtained by substituting the optimal power control policy in \eqref{opt:Pfinal} into \eqref{con:Srv} (instead of \eqref{eq:Pconst} with $P_0 = P_{\max}/W_{\max}$  as in Section \eqref{BLOnly}).

\begin{rem}
    \em{Since the power allocation in {\eqref{opt:Pfinal} and the bandwidth allocation found with stochastic optimization yield the unique solution that satisfies the necessary conditions,} the obtained solution is globally optimal to problem \eqref{prob:RA} in the
symmetric scenario.}
\end{rem}

\subsubsection{Optimizing $W_k$ and $P_k({\bf g})$ with Unsupervised Learning}
Even in the symmetric scenario, the optimal power allocation policy in \eqref{opt:Pfinal} is not in closed-form. In general asymmetric cases, the expression of $g_{k}^\mathrm{th}(\bm{g})$ cannot be derived from \eqref{opt:h}, and again there is no closed-form solution of $P_k(\bm{g})$. To avoid to use high complexity numerical method such as FEM to find the solution of the functional optimization for $P_k(\bm{g})$, we apply the method in Section \ref{subsec:funcDNN} to solve problem \eqref{prob:RA}, i.e., we turn to solving the following problem,
\begin{align}   \label{prob:RALag}
    \mathop \mathrm{max} \limits_{\lambda_k, h(\bm{g}), v_k(\bm{g})} \mathop \mathrm{min} \limits_{W_k, P_k(\bm{g})} &\  L_2 \\
    \text{s.t.} &\  P_k(\bm{g}) \!\geq\! 0, h(\bm{g}) \!\geq\! 0, \lambda_k \!\geq\! 0. \nonumber
\end{align}

We approximate $P_k(\bm{g})$ by $\mathcal{N}_P(\bm{g};\bm{\omega}_P)$, which is a DNN with model parameters $\bm{\omega}_P$, input $\bm{g}$, and output ${\left[\hat{P}_1(\bm{g};\bm{\omega}_P),\cdots\!,\hat{P}_K(\bm{g};\bm{\omega}_P)\right]}^\mathrm{T}$. In order not to lose any information during the forward propagation, the dimension of each hidden layer is set to be the number of users, which is the same as the input and output dimensions.
As mentioned before \eqref{opt:h}, the optimal solution of problem \eqref{prob:RA} is obtained when the equality in \eqref{con:Pmax} holds. By applying \texttt{Softmax} function as the activation function in the output layer, we can guarantee that $\hat{P}_k(\bm{g};\bm{\omega}_P) \geq 0$ and $\sum_{k=1}^K{\hat{P}_k(\bm{g};\bm{\omega}_P)}=P_{\max}$. Thereby, the term $\int_{\mathbb{R}_+^K} \!{\left[h(\bm{g}) \!\left(\sum_{k=1}^{K} {P_k(\bm{g})} \!-\! P_\mathrm{max}\!\right) - \sum_{k=1}^{K} {v_k(\bm{g}) P_k(\bm{g})}\right] \!\mathrm{d}\bm{g}}$ can be removed from the Lagrangian, and the corresponding Lagrange multiplier functions $h(\bm{g})$ and $v_k(\bm{g})$ can also be removed.
%
By replacing $P_k(\bm{g})$ in \eqref{prob:RALag} with $\hat{P}_k(\bm{g};\bm{\omega}_P)$, the joint power and bandwidth allocation optimization problem then becomes,
\begin{align}   \label{prob:RANN}
    \mathop \mathrm{max} \limits_{\lambda_k} \!\mathop \mathrm{min} \limits_{W_k, \bm{\omega}_P} \ & \hat{L}_2 \! \triangleq\! \sum_{k=1}^{K} \left[{W_k} \!+\! {\lambda_k \!\left( \mathbb{E}_{\bm{g}} \!\left\{\!e^{- \vartheta_k \hat{s}_k}\!\right\} \!-\! e^{-\vartheta_k B^\mathrm{E}_k} \right)}\right] \\
    \text{s.t.}  \ & \lambda_k \!\geq\! 0, \nonumber
\end{align}
where $\hat{s}_k \!=\! \frac{\tau W_k}{u \ln\!{2}} \left[\ln\!\left(\!1 \!+\! \frac{\alpha_k g_k \hat{P}_k(\bm{g};\bm{\omega}_P)}{N_0 W_k}\!\right) \!-\! \frac{Q_\mathrm{G}^{-1}\!\left({\varepsilon^\mathrm{c}_k}\right)} {\sqrt{\tau W_k}}\right]$.

The model parameters of the DNN $\bm{\omega}_P$, the allocated bandwidth $W_k$, $k=1,...,K$, and the Lagrange multipliers $\lambda_k$, $k=1,...,K$ can be obtained  from the following iterations,
\begin{align}
    \bm{\omega}_P^{(t+1)} &=\! \bm{\omega}_P^{(t)} \!-\! \phi_{\bm{\omega}_P}(t) \nabla_{\bm{\omega}_P} \hat{L}_2^{(t)} =\! \bm{\omega}_P^{(t)} \!-\! \phi_{\bm{\omega}_P}(t) P_\mathrm{max} \nabla_{\bm{\omega}_P} \mathcal{N}\left(\bm{g};\bm{\omega}_P^{(t)}\right) \nabla_{\hat{\bm{P}}} \hat{L}_2^{(t)}, \label{trn:Para} \\
    W_k^{(t+1)} &=\! {\left[W_k^{(t)} \!-\! \phi_W(t) \frac{\partial \hat{L}_2^{(t)}}{\partial W_k}\right]}^+, \label{trn:BW} \\
    \lambda_k^{(t+1)} &=\! {\left[\lambda_k^{(t)} \!+\! \phi_{\lambda}(t) \frac{\partial \hat{L}_2^{(t)}}{\partial \lambda_k}\right]}^+ =\! {\left[\lambda_k^{(t)} \!+\! \phi_{\lambda}(t) \frac{1}{N_\mathrm{b}} \sum_{n=1}^{N_\mathrm{b}}{\left( e^{-\! \vartheta_k \hat{s}_{k}^{(t,n)}} \!\!-\! e^{-\vartheta_k B^\mathrm{E}_k} \right)}\right]}^+, \label{trn:Lag}
\end{align}
where $\hat{L}_2^{(t)} \!\triangleq\! \frac{1}{N_\mathrm{b}} \sum_{n=1}^{N_\mathrm{b}}\sum_{k=1}^{K} {\left[W_k \!+\! \lambda_k \!\left( e^{- \vartheta_k \hat{s}_{k}^{(t,n)}} \!-\! e^{-\vartheta_k B^\mathrm{E}_k} \right)\right]}$, $\hat{s}_{k}^{(t,n)}$ is the $n$th realization of the achievable rate in the $t$th iteration, and $N_\mathrm{b}$ is number of realizations of small-scale channel gains in each iteration.
The gradient matrix of the DNN \emph{w.r.t.} the model parameters $\nabla_{\bm{\omega}_P} \mathcal{N}\left(\bm{g};\bm{\omega}_P^{(t)}\right)$ can be computed through backward propagation, and the gradient vector $\nabla_{\hat{\bm{P}}} \hat{L}_2^{(t)}$ is with the dimension of $K$ and the $k$th element of $- \frac{1}{N_\mathrm{b}} \sum_{n=1}^{N_\mathrm{b}} {\lambda_k^{(t)} \vartheta_k \frac{\partial \hat{s}_{k}^{(t,n)}}{\partial \hat{P}_k} e^{- \vartheta_k \hat{s}_{k}^{(t,n)}}}$.

\begin{rem} \label{rem:cnvrg}
\em{From \eqref{trn:Lag}, we can find that the iteration converges only if $\frac{1}{N_\mathrm{b}} \sum_{n=1}^{N_\mathrm{b}}{\left( e^{-\! \vartheta_k \hat{s}_{k}^{(t,n)}} \!\!-\! e^{-\vartheta_k B^\mathrm{E}_k} \right)}$ $\to 0$. This means that the QoS requirement in \eqref{con:Q} can be ensured when the iterations converge, because the constraints are judiciously controlled by the unsupervised learning framework.}
\end{rem}
\begin{rem}
\em{For mobile users, $\bm{\omega}_P$, $W_k$ and $\lambda_k$, $k=1,...,K$ need to be found again whenever their large-scale channel gains vary, say from the iterations in \eqref{trn:Para}, \eqref{trn:BW} and \eqref{trn:Lag} with random initial values. To accelerate convergence, we can employ pre-training, where the well-trained values of $\bm{\omega}_P$, $W_k$ and $\lambda_k$, $k=1,...,K$ with fixed user locations are used to initialize the iteration for re-training when the user locations change. Alternatively, we can also find the mapping from all environment parameters to the bandwidth and power allocation by further converting the variable optimization for $W_k$ into a functional optimization problem as in Section \ref{subsec:func}.}
\end{rem}

\section{Simulation Results}    \label{sec:Results}
In this section, we evaluate the performance achieved by the unsupervised deep learning when solving the variable and functional optimization problems in the previous section. For the bandwidth allocation problem without power allocation, we compare the performance of unsupervised learning with supervised learning in terms of approximation accuracy of the policy solution and the QoS violation. For the two-timescale bandwidth and power allocation problem, we compare the unsupervised learning with the global optimal solution in the symmetric scenario, considering that obtaining the labels for supervised learning is prohibitive.

We consider multiple users in a cell with radius of 250 m. At the beginning of  each slot, the small-scale channel gains of all the users are randomly generated from Rayleigh distribution. Other simulation parameters are listed in Table \ref{tab:SimParam}, {unless otherwise specified}.
\begin{table}[htbp]
	\small
	\renewcommand{\arraystretch}{1.3}
	\caption{Simulation Parameters}	\label{tab:SimParam}
	\begin{center}\vspace{-0.4cm}
	\begin{tabular}{|p{5.5cm}|p{3.2cm}|}
		\hline
 		Duration of each slot $T_\mathrm{s}$ & $0.1$~ms \\ \hline
		Duration of DL transmission  $\tau$ & $0.05$~ms \\ \hline
        Transmission delay $D^\mathrm{t}$ & $1$~slot ($0.1$~ms) \cite{Condoluci2017Reserv}\\ \hline
        Decoding delay $D^\mathrm{c}$ & $1$~slot ($0.1$~ms) \cite{Condoluci2017Reserv}\\ \hline
        Overall packet loss probability $\varepsilon_\mathrm{max}$ & $10^{-5}$ \\ \hline
        DL delay bound $D_\mathrm{max}$ & $10$~slots ($1$~ms) \\ \hline
        Maximal transmit power of BS $P_\mathrm{max}$ & $43$~dBm \\ \hline
		Path loss model $-10\lg(\alpha)$ & $35.3+37.6 \lg(d_k)$ \\ \hline
        Number of antennas $N_\mathrm{t}$ & 8 \\ \hline
		Single-sided noise spectral density $N_0$ & $-173$~dBm/Hz \\ \hline
		Packet size $u$ & $20$~bytes ($160$~bits) \cite{3GPP2016Scenarios} \\ \hline
        Average packet arrival rate $a$ & $0.2$~packets/slot \\ \hline
	\end{tabular}
	\end{center}
	\vspace{-0.7cm}
\end{table}

We apply fully-connected DNNs in learning algorithms, and use \texttt{TanH} in the input layer and the hidden layers as an example activation function, where similar results can be obtained with other activation functions. The activation functions for the output layers will be introduced later. The fine-tuned batch size for learning is $N_\mathrm{b} \!=\! 100$.

\subsection{Bandwidth Allocation without Power Allocation}
The users uniformly distributed along a road, which is with $50$~m minimal distance away from the BS. Since the bandwidth allocation without power allocation is independent for each user, we only consider the  bandwidth and QoS constraint of one user.

The two DNNs have six hidden layers, and each layer has $16$ neurons. We use \texttt{Softplus} in the output layers in all DNNs to ensure that the outputs are positive. The learning rate is $\phi(t) \!=\! 0.5/(1 \!+\! 10^{-4} t)$.
To evaluate the performance of learning in terms of the approximation accuracy to the optimal policy and the QoS guarantee, we define the relative error of the learnt bandwidth allocation to the optimal solution as $\sigma \!\triangleq\! \left|\hat{W}(\alpha)/W^*(\alpha) \!-\! 1\right|$, and the QoS violation of the learnt solution as $\nu \!\triangleq\! \left(\mathbb{E}_{g} \left\{e^{\vartheta (B^\mathrm{E} - \hat{s})}\right\} \!-\! 1 \right)^+ $.

\begin{figure}[ht]
        \centering
        \begin{minipage}[t]{0.5\textwidth}
        \includegraphics[width=1\textwidth]{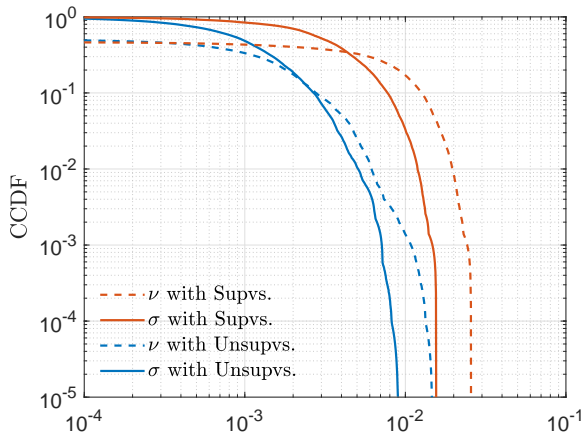}
        \end{minipage}\vspace{-0.2cm}
        \caption{Complementary cumulative distributions of the relative approximation error of $\hat W$ and the QoS violation.}
        \label{fig:Conv}
        \vspace{-0.2cm}
\end{figure}

In Fig. \ref{fig:Conv}, we show the complementary cumulative distributions (CCDF) of $\sigma$ and $\nu$ achieved by the unsupervised learning approach in Section \ref{Unsurp-1}, and those obtained by supervised learning approach where a DNN is used to learn the optimal policy $W^*(\alpha)$ and is trained by taking the optimal solutions of problem in \eqref{eq:SUW} as labels. The results are obtained through $100$ trails, where in each trail the NNs are trained through $10\,000$ iterations and are tested on $1\,000$ realizations of the large-scale channel gains.
It is shown that the unsupervised learning approach outperforms the supervised learning approach. With unsupervised learning, the relative approximation error of the allocated bandwidth is less than $1$\%  and the QoS violation probability is less than $2$\% with a probability of $99.999\%$.

\vspace{-2mm}
\subsection{Joint Bandwidth and Power Allocation}
To show the performance gap of the solution obtained with unsupervised learning from the global optimal solution, we first consider a symmetric scenario, where all users are in the cell-edge, i.e., the user-BS distances are $250$~m. Then, we evaluate the performance considering an asymmetric scenario, where the users are uniformly located in the road with $50$~m minimal distance away from the BS, i.e., the user-BS distances are distributed from 50 m to $250$~m.
The DNN has two hidden layers, and the number of neurons in each layer equals to the number of users. We use \texttt{Softmax} in the output layer to ensure the maximum transmit power constraint and $P_k({\bm g}) \!\geq\! 0$ for all $k \!=\! 1,\cdots\!,K$. The learning rate is set to be $\phi(t) \!=\! 1/(1 \!+\! 0.1 t)$, which turns out to be a good setting according to our experience.

The joint optimal policy (with legend ``w MUD w FD")  is obtained from the method in Section \ref{sec:Sym} with around $200$ iterations in the symmetric scenario, which exploits multi-user diversity by dynamically adjusting the transmit power according the small-scale channel gains of users, and exploits frequency diversity by frequency hopping.
The learning-based bandwidth and power allocation policy (with legend ``w MUD w FD (NN)") is obtained from the iterations in \eqref{trn:Para}, \eqref{trn:BW} and \eqref{trn:Lag} with random initial values. In each slot, the channel realizations in recent $N_\mathrm{b}$ slots are taken as a batch, which is used for $10$ iterations in each slot. The training procedure converges after $100$ slots, unless otherwise specified.

To show the gain from multi-user diversity, we compare the joint optimal policy with the optimal bandwidth allocation policy obtained through \eqref{opt:W} after sufficient iterations, where the transmit power is equally allocated in the frequency domain without exploiting multi-user diversity (with legend ``w/o MUD w FD''). To show the gain from frequency diversity, we compare with a heuristic policy in \cite{Chengjian2017GCw}, which also exploits multi-user diversity by scheduling the users according to their small-scale channel gains but does not exploit frequency diversity (with legend ``w MUD w/o FD"). Finally, we show the performance of the policy in \cite{She2018CrossLayer} as a baseline, which optimizes the bandwidth allocation, but exploits neither multi-user diversity nor frequency diversity (with legend ``w/o MUD w/o FD").


\begin{figure}[htbp]
	\vspace{0.05cm}
	\centering
	\begin{minipage}[t]{0.6\textwidth}
	\subfigure[{Symmetric scenario.}]{\label{fig:BW_Sym}
		\includegraphics[width=0.4936\textwidth]{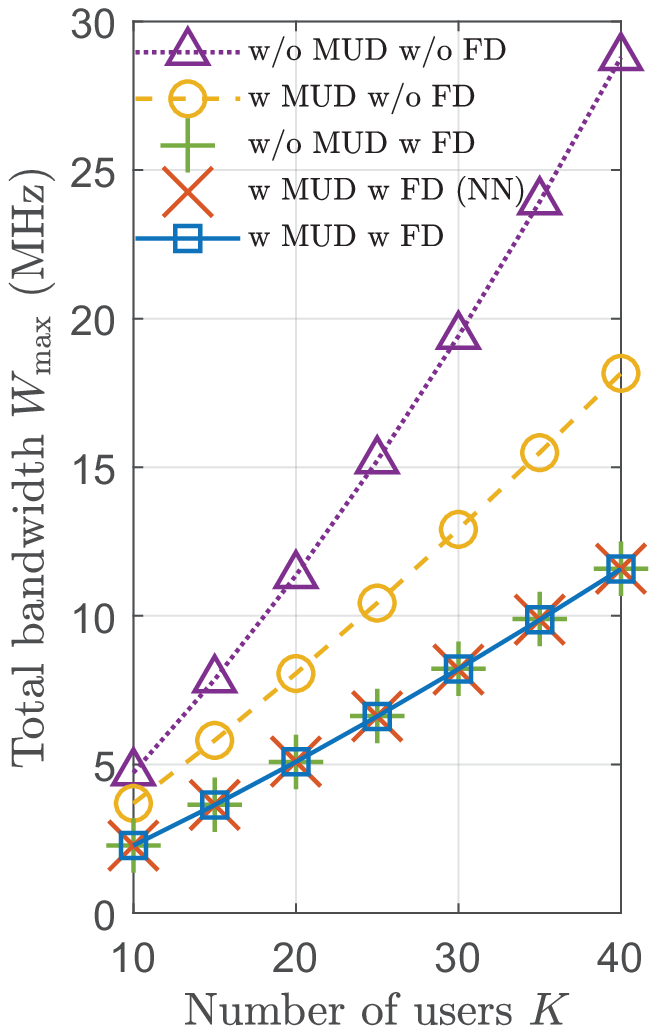}} 
	\subfigure[{Asymmetric scenario.}]{\label{fig:BW_Asym}
		\includegraphics[width=0.4664\textwidth]{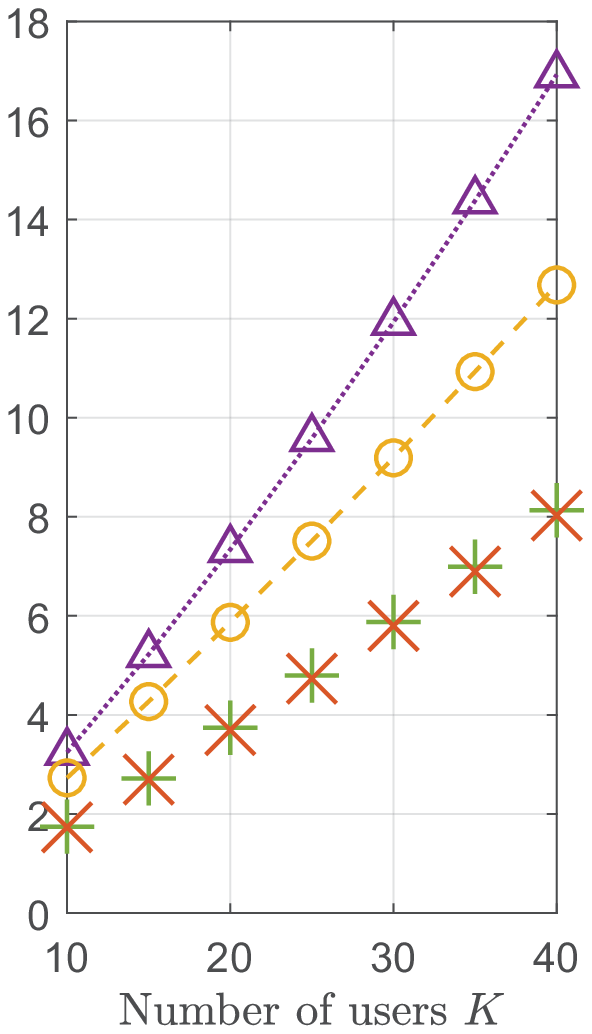}} 
	\end{minipage}
    \vspace{-0.3cm}
    \caption{Total bandwidth required to support the QoS of each user.}
	\vspace{-0.2cm}
\end{figure}

 In Fig. \ref{fig:BW_Sym}, we provide the results in the symmetric scenario. It shows that the performance of learning-based policy (i.e., ``w MUD w FD (NN)") is almost the same as the global optimal policy derived in \eqref{opt:Pfinal} (i.e., ``w MUD w FD"). In Fig. \ref{fig:BW_Asym}, we provide the results in the asymmetric scenario, where only the learning-based policy is simulated since the optimal solution is not available in this scenario. From both scenarios we can see that exploiting multi-user diversity or frequency diversity individually can significantly improve the bandwidth efficiency, while the gain from frequency diversity is larger. Once the frequency diversity is exploited, multi-user diversity only provides marginal performance gain.

To show the convergency of the learning-based solution, we consider the sum of the absolute values of average gradients
$\zeta^{(t)} \!\triangleq\! {\left\|{\mathbb{E}_{\bm{g}} \!\left\{\nabla_{\bm{\omega}_P} \hat{L}^{(t)}\!\right\}}\right\|}_1 \!+\! \sum_{k=1}^{K} \!{\left|\mathbb{E}_{\bm{g}} \!\left\{\frac{\partial \hat{L}^{(t)}}{\partial W_k}\!\right\}\!\right|} \!+\! \sum_{k=1}^{K} \!{\left|\mathbb{E}_{\bm{g}} \!\left\{\frac{\partial \hat{L}^{(t)}}{\partial \lambda_k}\!\right\}\!\right|}$
and the QoS constraint violation
$\xi^{(t)} \!\triangleq\! \sum_{k=1}^{K} \!{{\left[\mathbb{E}_{\bm{g}} \!\left\{e^{\vartheta_k \left(B^\mathrm{E}_k - \hat{s}_k^{(t)}\right)}\!\!\right\} \!-\! 1\right]}^+} \!\!\Big/\! K$.
The training algorithm in \eqref{trn:Para}, \eqref{trn:BW} and \eqref{trn:Lag} is considered to be converged at the $t$th slot if $\zeta^{(t)} \!<\! 1\%$ and $\xi^{(t)} \!<\! 1\%$.

\begin{table}[htbp]
	\small
	\renewcommand{\arraystretch}{1.3}
	\caption{Number of Time Slots for Convergence, Asymmetric scenario, $T_\mathrm{s}=0.1$~milliseconds}	\label{tab:CvgRate}
	\begin{center} \vspace{-0.4cm}
    \begin{tabular}{|c|c|c|}
    \hline
    Convergence percentage & $99.9\%$ & $99.99\%$ \\ \hline
    w/o pre-training & $5\,000$ & $>\!10\,000$ \\ \hline
    w pre-training & $3$ & $1\,000$ \\ \hline
    \end{tabular}
    \end{center}
	\vspace{-0.4cm}
\end{table}

The convergence speeds with and without pre-training are shown in Table \ref{tab:CvgRate}, which are obtained from $100\,000$ trails. For the results without pre-training, $40$ users are randomly dropped in the road in each trail and the realizations of their large- and small-scale channel gains are used to train $\bm{\omega}_P$, $W_k$ and $\lambda_k$, $k \!=\! 1,\cdots\!,K$, with random initializations. For the results with pre-training, all users move at the velocity of $72$~kph along the road in the same direction. The well-trained values of $\bm{\omega}_P$, $W_k$ and $\lambda_k$, $k \!=\! 1,\cdots\!,K$, are fine-tuned every $0.1$~s using the channels at the new locations. Without pre-training, $10\,000$ time slots (i.e., 1 s) are required to achieve $99.99\%$ convergence percentage, i.e., the QoS of each user is ensured with a probability of $99.99\%$ according to Remark~\ref{rem:cnvrg}.
We can see that the pre-training, which can be accomplished off-line, shortens the convergence time significantly. The complexity of the training is low. A computer with {Intel\textregistered}\ {Core\texttrademark}\ {i7-6700} CPU is able to finish around $1\,000$ iterations in $0.1$~s without using the acceleration from GPU.

\section{Conclusion}
In this paper, we proved that the problem of finding the mapping from environment parameters to the solutions of constrained variable optimizations can be formulated as functional optimizations with instantaneous constraints, and established a unified unsupervised deep learning framework to solve functional optimizations with both instantaneous and statistic constraints. We considered two example problems in downlink URLLC to illustrate how to apply this framework. The first problem is variable optimization, where bandwidth allocation is optimized according to large-scale channel gains. The second problem is a hybrid variable and functional optimization with two types of constraints, where we jointly optimized bandwidth allocation according to large-scale channel gains and power allocation according to small-scale channel gains.
Simulations results showed that, for the bandwidth allocation problem, unsupervised learning is superior to the supervised learning in both the accuracy of approximating the optimal solution and the guarantee of the QoS constraint. For the joint bandwidth allocation and power allocation problem, the learning-based solution performs almost the same as the global optimal solution in a symmetric scenario. For both problems, the QoS achieved by the solution using unsupervised learning can be guaranteed with very high probability. The training algorithm converges rapidly with pre-training, and is with low computational complexity. As a byproduct, the optimization results also showed that the bandwidth utilization efficiency of URLLC can be improved more significantly by exploiting frequency diversity than by multi-user diversity.

\appendices
\section{Proof of Proposition 1}
\label{App:Upgrade}
\renewcommand{\theequation}{A.\arabic{equation}}
\setcounter{equation}{0}

\begin{proof}
We first prove that $\bm{x}^*(\bm{\theta})$ is optimal for problem \eqref{prob:FuncOpt}.
Denote $\bm{x}^* (\bm{\theta}_1)$ as an optimal solution of problem \eqref{prob:VarOpt} given an arbitrary realization $\bm{\theta}_1 \!\in\! \mathcal{D}_\theta$, and denote the objective function in problem \eqref{prob:FuncOpt} as $\mathcal{F} \left[\bm{x}(\bm{\theta})\right] \!\triangleq\! \int_{\bm{\theta} \in \mathcal{D}_\theta} {f\left[\bm{x}(\bm{\theta});\bm{\theta}\right] p(\bm{\theta}) \mathrm{d} \bm{\theta}}$.
Let $\bm{x}_1 (\bm{\theta}), \bm{\theta} \!\in\! \mathcal{D}_\theta$ be an arbitrary feasible solution of problem \eqref{prob:FuncOpt}. Since problems \eqref{prob:VarOpt} and \eqref{prob:FuncOpt} have the same constraints, they have the same feasible region. Thus, $\bm{x}_1 (\bm{\theta}_1)$ is a feasible solution of problem \eqref{prob:VarOpt}. Given the realization $\bm{\theta}_1$, the optimal solution of problem \eqref{prob:VarOpt} is better than any feasible solutions of problem \eqref{prob:VarOpt}, i.e.,
\begin{align}
    f\left[\bm{x}^* (\bm{\theta}_1);\bm{\theta}_1\right] - f\left[\bm{x}_1 (\bm{\theta}_1);\bm{\theta}_1\right] \leq 0, \  \forall \bm{\theta}_1 \in \mathcal{D}_\theta.
\end{align}
Since $p(\bm{\theta}) \!\geq\! 0$, we further have,
\begin{align}\label{eq:xstar}
    \mathcal{F} \left[\bm{x}^* (\bm{\theta})\right] - \mathcal{F} \left[\bm{x}_0 (\bm{\theta})\right] = \int_{\bm{\theta} \in \mathcal{D}_\theta} {\left[f\left[\bm{x}^* (\bm{\theta});\bm{\theta}\right] - f\left[\bm{x}_0 (\bm{\theta});\bm{\theta}\right]\right] p(\bm{\theta}) \mathrm{d} \bm{\theta}} \leq 0.
\end{align}
Since $\bm{x}^* (\bm{\theta}), \bm{\theta} \in \mathcal{D}_\theta,$ satisfies all the constraints in problem \eqref{prob:FuncOpt}, it is a feasible solution of problem \eqref{prob:FuncOpt}. \eqref{eq:xstar} indicates that $\bm{x}^* (\bm{\theta}), \bm{\theta} \in \mathcal{D}_\theta$ is better than an arbitrary solution of problem \eqref{prob:FuncOpt}. Thus, it is optimal for problem \eqref{prob:FuncOpt}.
%

In what follows, we prove that the value of $\bm{x}_{\rm opt}(\bm{\theta})$ for arbitrary realization of $\bm{\theta}$ is optimal for problem \eqref{prob:VarOpt} with probability one. From the definition of $\bm{x}_{\rm opt} (\bm{\theta})$ and $\bm{x}^*(\bm{\theta})$, we have
\begin{align}   \label{eq:theta}
    f\left[\bm{x}_{\rm opt} (\bm{\theta});\bm{\theta}\right] - f\left[\bm{x}^*(\bm{\theta}),\bm{\theta}\right] \geq 0, \  \forall \bm{\theta} \in \mathcal{D}_\theta.
\end{align}
Suppose there exists a non-zero measure set, $\mathcal{D}^+_\theta$, such that for any $\bm{\theta}' \in \mathcal{D}^+_\theta$, $\bm{x}_{\rm opt} (\bm{\theta}')$ is not optimal for problem \eqref{prob:VarOpt}. In other words, there exists a $\delta_0$ such that
\begin{align}\label{eq:delta0}
     f\left[\bm{x}_{\rm opt} (\bm{\theta}');\bm{\theta}'\right] - f\left[\bm{x}^*(\bm{\theta}'),\bm{\theta}'\right] \geq \delta_0>0, \forall \bm{\theta}' \in \mathcal{D}^+_\theta.
\end{align}
Here, a non-zero measure set is a set that $\Pr\{\bm{\theta}' \in \mathcal{D}^+_\theta\} > 0$.

From \eqref{eq:theta} and \eqref{eq:delta0}, we can derive that
\begin{align}
&\mathcal{F} \left[\bm{x}_{\rm opt} (\bm{\theta})\right] - \mathcal{F} \left[\bm{x}^* (\bm{\theta})\right]\nonumber\\
&=\int_{\bm{\theta} \in \mathcal{D}_\theta}{\left[f\left[\bm{x}_{\rm opt} (\bm{\theta});\bm{\theta}\right] - f\left[\bm{x}^*(\bm{\theta}),\bm{\theta}\right]\right]p(\bm{\theta}) \mathrm{d} \bm{\theta}}\nonumber\\
&\geq \int_{\bm{\theta} \in \mathcal{D}^+_\theta} {\left[f\left[\bm{x}_{\rm opt} (\bm{\theta});\bm{\theta}\right] - f\left[\bm{x}^*(\bm{\theta}),\bm{\theta}\right]\right] p(\bm{\theta}) \mathrm{d} \bm{\theta}} \nonumber\\
&\geq \int_{\bm{\theta} \in \mathcal{D}^+_\theta} {\delta_0 p(\bm{\theta}) \mathrm{d} \bm{\theta}} \nonumber\\
&= \delta_0 \Pr\{\bm{\theta} \in \mathcal{D}^+_\theta\} > 0.
\end{align}
This contradicts with the fact that $\bm{x}_{\rm opt} (\bm{\theta})$ is the optimal solution of problem \eqref{prob:FuncOpt}.

This completes the proof.
\end{proof}

\section{The method to compute \eqref{eq:DLx}, \eqref{eq:DLlamda} and \eqref{eq:DLnu}}
\label{App:derive}
\renewcommand{\theequation}{B.\arabic{equation}}
\setcounter{equation}{0}
\begin{proof}
For notational simplicity, we omitted the index of iteration $t$ in this appendix. To compute \eqref{eq:DLx}, \eqref{eq:DLlamda} and \eqref{eq:DLnu}, we only need to compute ${\nabla _{{\omega_{x}}}}{{\hat L}}$, ${\nabla _{{\omega_{{v}}}}}{{\hat L}}$ and $\partial{\hat L}/\partial{{\lambda}_j}$.

The value of ${\nabla _{{\omega_{x}}}}{{\hat L}}$ can be obtained from the following expression,
\begin{align}
{\nabla _{{\omega_{x}}}}{{\hat L}}
& = \int\limits_{{{\bm{\theta}}} \in {D_{{\theta}}}} { [{{\nabla _{{\omega_{x}}}}{\bm{\hat{x}}({\bm{\theta}})}}] \left[ {{\nabla _{{\bm{\hat{x}}({\bm{\theta}})}}}f\left( {\bm{\hat{x}}({\bm{\theta}})};{{\bm{\theta}}} \right)} \right]{p}\left( {{\bm{\theta}}} \right){\rm{d}}{{\bm{\theta}}}} \nonumber\\
&+\sum\limits_{j = I+1}^J {{\lambda}_j {\int\limits_{{{\bm{\theta}}} \in {D_{{\theta}}}} { [{{\nabla _{{\omega_{x}}}}{\bm{\hat{x}}({\bm{\theta}})}}] \left[ {{\nabla _{{\bm{\hat{x}}({\bm{\theta}})}}}C_j\left( \bm{\hat{x}}({\bm{\theta}});{{\bm{\theta}}} \right)} \right]{p}\left( {{\bm{\theta}}} \right){\rm{d}}{{\bm{\theta}}}} } } \nonumber\\
&+\sum\limits_{i = 1}^I {\int\limits_{{{\bm{\theta}}} \in {D_{{\theta}}}} {\hat{\lambda}_j\left( {{{\bm{\theta}}}} \right) [{{\nabla _{{\omega_{x}}}}{\bm{\hat{x}}({\bm{\theta}})}}] \left[ {{\nabla _{{\bm{\hat{x}}({\bm{\theta}})}}} C_i \left( {{\bm{\hat{x}}({\bm{\theta}})};{{\bm{\theta}}}} \right)} \right]{\rm{d}}{{\bm{\theta}}}} }, \label{eq:DLxB}
\end{align}
where ${{\nabla _{{\omega_{x}}}}{\bm{\hat{x}}({\bm{\theta}})}} \triangleq [{{\nabla _{{\omega_{x}}}}{\hat{x}_1({\bm{\theta}})}},...,{{\nabla _{{\omega_{x}}}}{\hat{x}_{N_x}({\bm{\theta}})}}]$ can be obtained via backward propagation.

The values of ${\nabla _{{\omega_{{v}}}}}{{\hat L}}$ and $\partial{\hat L}/\partial{{\lambda}_j}$ can be obtained from
\begin{align}
&{\nabla _{{\omega_{{v}}}}}{{\hat L}} = \sum\limits_{i = 1}^I {\int\limits_{{{\bm{\theta}}} \in {D_{{\theta}}}} {[{\nabla _{{\omega_{{v}}}}}\hat{{v}_i}({\bm{\theta}})]\left[ {C_i\left( {{\bm{\hat{x}}({\bm{\theta}})};{{\bm{\theta}}}} \right)} \right]{\rm{d}}{{\bm{\theta}}}} } \label{eq:DLlambdaB},\\
&\frac{\partial{\hat L}}{\partial{{\lambda}_j}} ={\int\limits_{{{\bm{\theta}}} \in {D_{{\theta}}}} {C_j\left( {{\bm{\hat{x}}({\bm{\theta}})};{{\bm{\theta}}}} \right){p}\left( {{\bm{\theta}}} \right){\rm{d}}{{\bm{\theta}}}}}\label{eq:DLnuB},
\end{align}
where ${{\nabla _{{\omega_{\lambda}}}}{\hat{\lambda_i}(\bm{\theta})}}, i=1,...,I$, can be obtained via backward propagation.
\end{proof}

\bibliographystyle{IEEEtran}
\bibliography{ref}

\end{document}